\documentclass[aps,groupedaddress,twocolumn,floatfix,pre]{revtex4}
\usepackage[colorlinks=true]{hyperref} 
\hypersetup{urlcolor=blue,linkcolor=red,citecolor=blue,colorlinks=true} 

\usepackage{ifthen}

\usepackage{fancyhdr}

\usepackage{graphics,epsfig}
\usepackage{amssymb,amsfonts}
\usepackage{amsmath}
\usepackage{mathtools}
\usepackage{color}
\usepackage{tikz}
\usepackage{lipsum}  
 \usepackage[utf8]{inputenc}
\usepackage[braket, qm]{qcircuit}

\usepackage{amsmath,amssymb,braket}

\usetikzlibrary{graphs,graphs.standard}

\usetikzlibrary{lindenmayersystems}

\usepackage[export]{adjustbox}
\newcommand{\eref}[1]{Eq.~(\ref{#1})}
\newcommand{\efig}[1]{Fig.~\ref{#1}}

\fancypagestyle{FirstPage}{
\lfoot{Thales Alenia Space Limited Distribution} 
}

\begin{document}
\title{Tensor network methods for quantum-inspired image processing and classical optics}

\author{Nicolas Allegra}
\email{nicolas.allegra@gmail.com}
\affiliation{Thales Alenia Space, Cannes}

\begin{abstract}
Tensor network methods strike a middle ground between fully-fledged quantum computing and classical computing, as they take inspiration from quantum systems to significantly speed up certain classical operations. Their strength lies in their compressive power and the wide variety of efficient algorithms that operate within this compressed space.
In this work, we focus on applying these methods to fundamental problems in image compression and processing and classical optics such as wave-front propagation and optical image formation, by using directly or indirectly parallels with quantum mechanics and computation. These quantum-inspired methods are expected to yield faster algorithms with applications ranging from astronomy and earth observation to microscopy and classical imaging more broadly.

\end{abstract}

\date{\today}

\maketitle
\section{Introduction }

For decades, scientists have tried to simulate quantum systems on classical computers. The challenge lies in the exponential memory needed, making direct simulations unfeasible for anything beyond very small systems or idealized systems. To overcome this, researchers had spent decades developing clever approximations for quantum and statistical lattice systems.
Although tensor networks—particularly matrix product states—can be traced back to R.J Baxter and his work on the corner transfer matrix method for solving $2d$ statistical mechanics problems \cite{baxter1968dimers}, their popular emergence occurred decades later in the nineties, with the major breakthrough of S.R. White’s density matrix renormalization group algorithm \cite{white1992density}, which compress quantum systems by retaining only the most relevant correlations. 

Recently, the field of tensor networks has seen a surge in interest due to their connections to quantum computing. Quantum researchers use tensor networks to design and simulate their quantum machines, while tensor networks experts borrow ideas from quantum theory to emulate and compete with quantum performance \cite{huang2019simulating,oh2023tensor,liu2023simulating,tindall2024efficient,patra2024efficient}. This feedback loop benefits both sides and benefits also scientists and engineers tackling complex real-world problems. To name a few examples, tensor networks have enabled immediate applications in machine learning \cite{stoudenmire2016supervised,glasser2018supervised}, fluid mechanics \cite{peddinti2024quantum,arenstein2025fast,van2025quantum,gourianov2025tensor,holscher2025quantum}, finance  \cite{rihito2025learning}, electromagnetism \cite{lively2025quantum}, material science \cite{hauck2024sfft} and computational chemistry \cite{jolly2025tensorized}. 
 
One of the first application of tensor networks outside of physics was an interesting study on image compression using matrix product states \cite{latorre2005image}. Although this 20-year-old paper did not have a significant impact on the field of image processing at the time, the rapidly growing field of quantum machine learning has recently revived interest in this topic \cite{dilip2022data,jobst2024efficient}. In the present work, we build upon these earlier studies by applying some of these quantum computing and tensor network compression ideas to image processing and Fourier optics.

\textbf{Outline of the paper  } 
In the first half of this paper, we focus on natural images and quantum-inspired methods for encoding them into quantum-like objects, which will prove useful for many applications.
Several types of matrix product encodings will be compared with one another, along with alternative tensor network structures such as tree tensor networks. \textcolor{black}{We will discuss their compression capabilities and explain their performances in terms of locality, scale separation and entropy using the language of statistical mechanics and quantum physics.}  The discussion will then shift to classical operations on images represented as tensor network operators, and how these can be applied to classical imaging and image processing. 
One of the goals of this paper is to draw parallels between Fourier optics and quantum mechanics, and to use these parallels to construct efficient approximations of classical optical propagation in very large dimensions by finding local Hamiltonians whose evolution represents these classical operations. These connections will be often made using ideas stemming from quantum computation—even though no quantum computers are involved—showing how quantum mechanics concepts can lead to efficient classical computations. 

\section{Natural images, resolution and large-scale limit}

We define a continuous image (sometimes called a scene or an objet) as the intensity $h$ of an electromagnetic field on a $2d$ plane. From this function of continuous variables, we can consider an approximation on a $2d$ grid  of size $L\times L=2^n\times 2^n$ by evaluating it at discrete points $(x,y)=\Big(\frac{a}{2^n},\frac{b}{2^n}\Big)$ with $a,b\in \{0,1,2,...,2^n-1\}$.
\begin{figure}[ht!]
\includegraphics[scale=0.3]{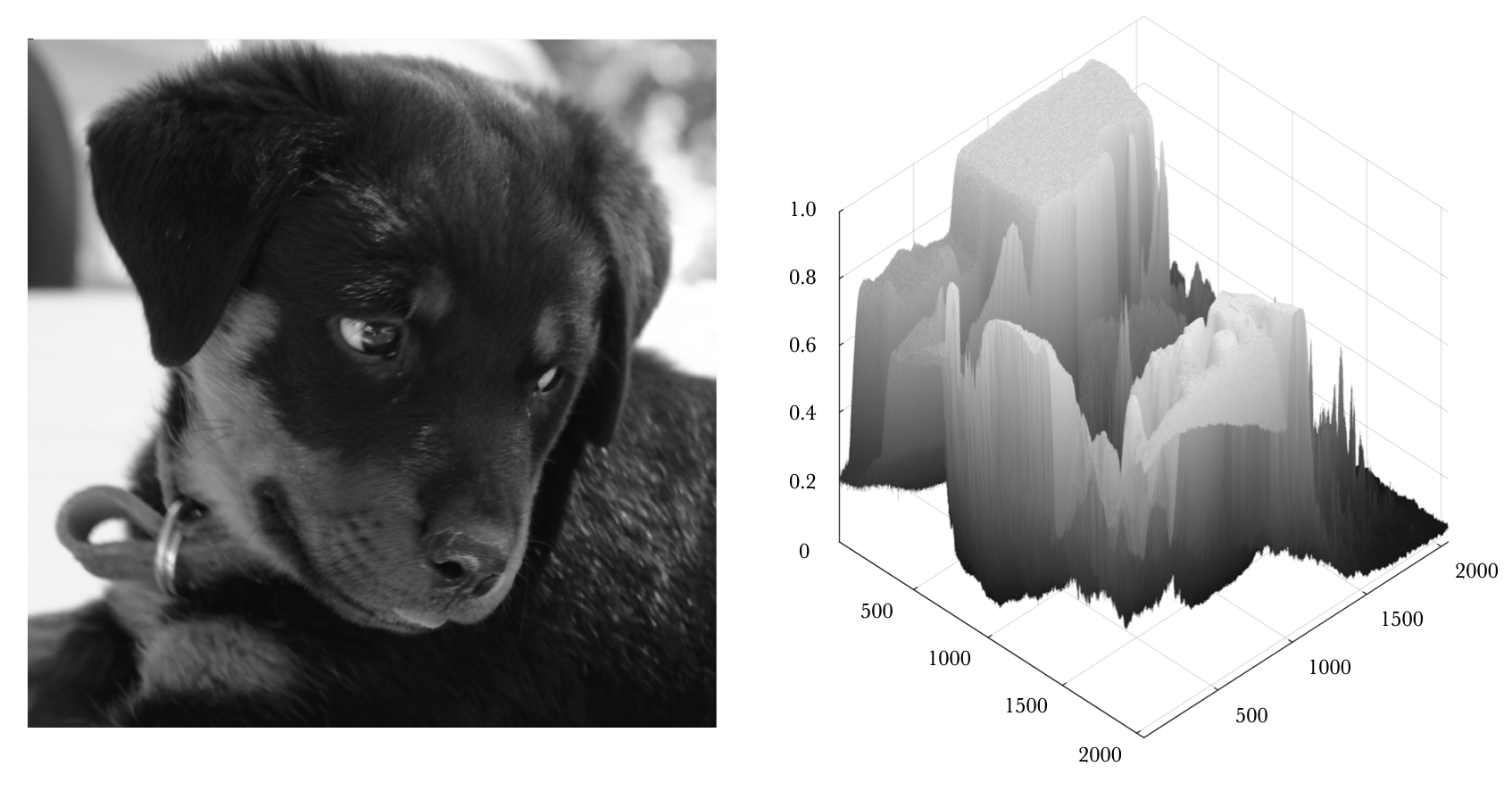} 
	\caption{A black and white picture and its height field representation. This picture will be used throughout this paper.}
		\label{fig_plane}
\end{figure}
As we can not get access to a real continuous image of the physical world (or a model of it), we will have to settle for a digital image as a proxy to the real world and try to separate what comes from the physical scene to what comes from the measurement device (the digital camera here). \textcolor{black}{The question we are interested in here, is how much information and correlation is relevant in an image at the continuous limit and how this can be used to construct an effective model that compress and approximate efficiently the image as well as implement faster optical and imaging simulations on classical computers.}

Natural images are closer to real physical systems than they are from random samples of a generic distribution, moreover they are known to be scale invariant with a power spectrum that obeys a power-law distribution \cite{ruderman1994statistics,ruderman1997origins}. In a more recent study \cite{jobst2024efficient}, the authors showed that, over a large dataset of various natural images of various shapes and forms, the behavior of the Fourier amplitudes follows $\langle \hat{h}(k_x,k_y)\rangle\sim\mathcal{O}(k_x^{-1.1})\mathcal{O}(k_y^{-1.07})$. In this present work, we choose to work with a single \textquotedblleft statistically average\textquotedblright image \footnotemark\footnotetext{As a remark, although the use of \textquotedblleft Lena\textquotedblright
 as the go-to image in the field of image processing has, understandably so, fallen out of favor in recent years for many reasons, its non-power-law behavior could be added to the list of grievances.} (\textit{cf.} \efig{fig_plane}) that is very representative of the power-law behavior of natural images. The picture showed numerous times in this paper \footnotemark\footnotetext{Original image: Roxy.JPEG, dimension $3264 \times 2448$, taken with a SONY DSC-F828 camera, sRGB IEC61966-2.1, focal length 38.4~mm, f-number $f/3.2$, exposure time $1/60$~s.}  follows this behavior honorably well (it is actually closer to $\hat{h}(k)\sim k^{-1}$) and saves us from the time/resource consuming task of averaging over a large dataset of images, with all the generalization and statistical limitations that it entails.  

\begin{figure}[ht!]
\includegraphics[scale=0.25]{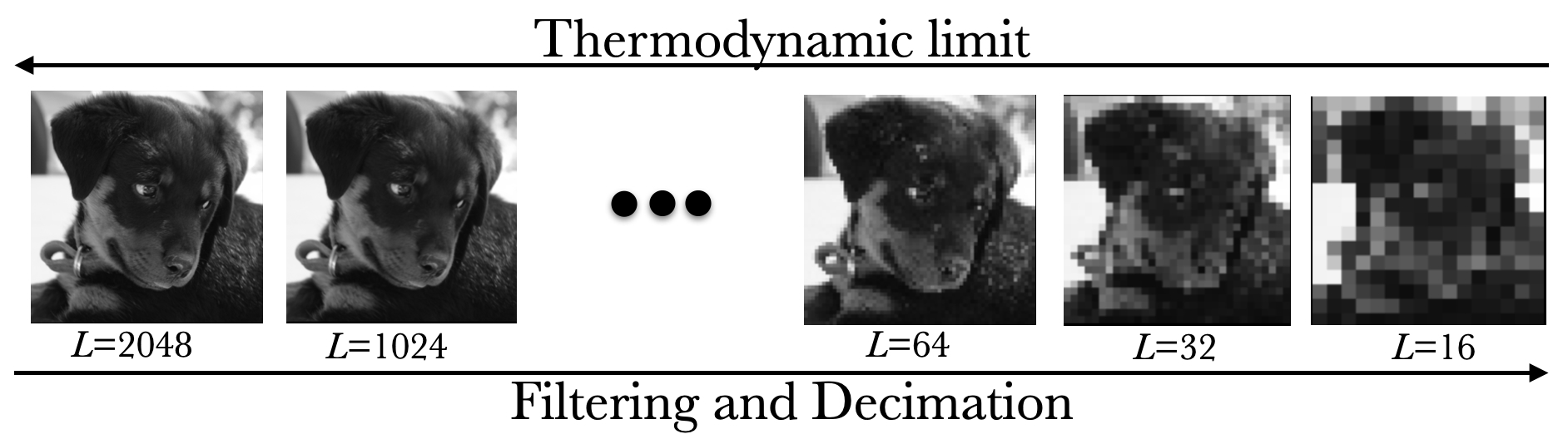} 
	\caption{Physical interpretation of an image at various resolutions. The downsampling algorithm does not introduce RG relevant features.}
		\label{fig_scale}
\end{figure}

\textcolor{black}{
We construct a multi-scale representation of a high-resolution image by iteratively downsampling by a factor of two, starting from a large image of size $L \times L$ down to coarse resolutions \efig{fig_scale}. The downsampling is performed using the \texttt{imresize} routine from the \texttt{ImageTransformations.jl} package. Since the procedure involves only coarse-graining operations and no upsampling, it does not introduce spurious degrees of freedom. This argument has been checked numerically, since all tested resizing algorithms have lead to the same results. Each step effectively integrates out short-wavelength fluctuations while preserving long-wavelength structure, in close analogy with a real-space renormalization group (RG) transformation. For natural images, and with standard interpolation, this transformation is well approximated by a low-pass filtering followed by decimation.}

\textcolor{black}{
 From an information-theoretic perspective, the entropy of the image decreases monotonically along this flow, while mutual information between well-separated regions is preserved up to cutoff-scale corrections. Any resampling-induced artifacts are confined to the ultraviolet scale and correspond to RG-irrelevant perturbations that do not affect the large-scale behavior \footnotemark\footnotetext{Differences between interpolation kernels act as RG-irrelevant operators and are washed out under further coarse-graining. Let us notice that the opposite process of starting with a small image would not satisfied the requirements.}. As a result, the sequence of resized images may be regarded as multiple realizations of the same effective system at different linear sizes $L$, allowing one to analyze scale-dependent properties (such as entropy scaling, correlation structure, or compression efficiency) using the standard tools of statistical mechanics in the large-$L$ limit.}

 \textcolor{black}{
The down-sampling procedure used in this work is deterministic, and therefore yields a single representative image at each resolution rather than a full statistical ensemble. As a result, the physical interpretation presented above should be understood in an effective or self-averaging sense, appropriate for large natural images. To make this analogy more explicit, an ensemble at fixed resolution could be generated by introducing randomness only at the ultraviolet scale (for instance by randomizing sub-pixel offsets or local averaging kernels during the coarse-graining step) while preserving long-wavelength structure. Such randomized coarse-graining procedures are analogous to stochastic block-spin renormalization and differ only by RG-irrelevant details. While we do not follow this route in the present work, it represents a
necessary path toward a more thorough statistical analysis and will be
explored in future research.}

\section{Quantum encodings of images}

\begin{figure}[ht!]
\includegraphics[scale=0.16]{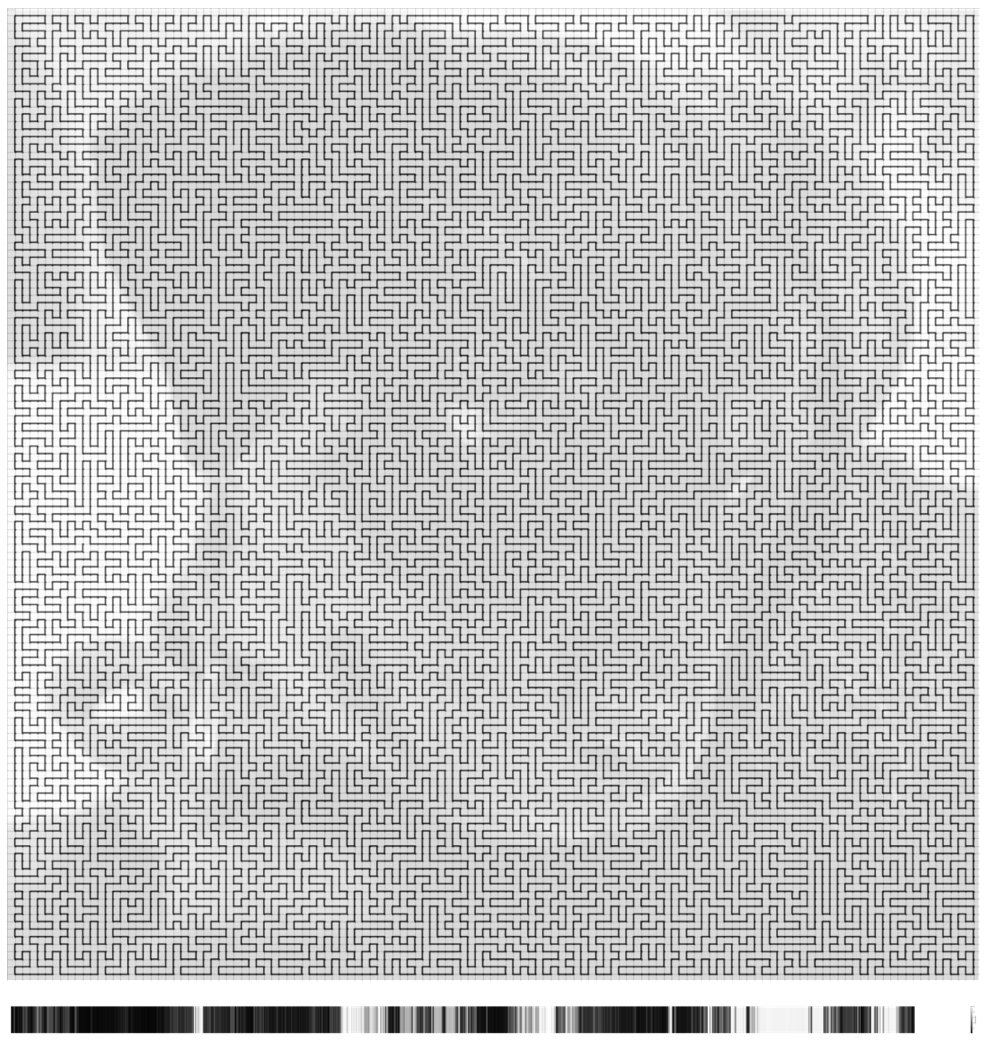} 
	\caption{Illustration of matrix to vector reshaping used in FRQI \eref{FRQI}. Below the image, the corresponding vector noted $x_i$ with $i=0...2^{2n-1}$.}
	\label{fig_SARW}
\end{figure}

Next, we are mapping this matrix to a higher dimensional object (a tensor) and then operate compression within this space. The first tensor image encoding that we are considering comes from the field of quantum machine learning, where researchers are interested in efficient ways to feed data to quantum computers. The so-called Flexible Representation of Quantum Images (FRQI) \cite{le2011flexible} is a quantum encoding scheme that non-trivially represents images by a multi-qubit quantum state $|h\rangle$ by factorizing the pixel values and their positions.  More precisely, an image of size $2^n \times 2^n$ is encoded using a normalized superposition of $2n+1$ qubits   
\begin{equation}\label{FRQI}
|\psi\rangle=\frac{1}{2^n}\sum_{i=0}^{2^{2n}-1} |c(x_i)\rangle\otimes |i\rangle,
\end{equation}
with $|c(x_i)\rangle=\cos\big(\frac{\pi}{2}x_i\big) |0\rangle + \sin\big(\frac{\pi}{2} x_i\big) |1\rangle$\footnotemark\footnotetext{Other kernels are envisageable and well-studied in quantum machine learning, depending on data, context and purposes.}, where $x_i$ encodes the intensity of the pixel at position $i$ (see \efig{fig_SARW}), and the basis state $|i\rangle$ represents the pixel's position in the image.

\begin{figure}[ht!]
\includegraphics[scale=0.25]{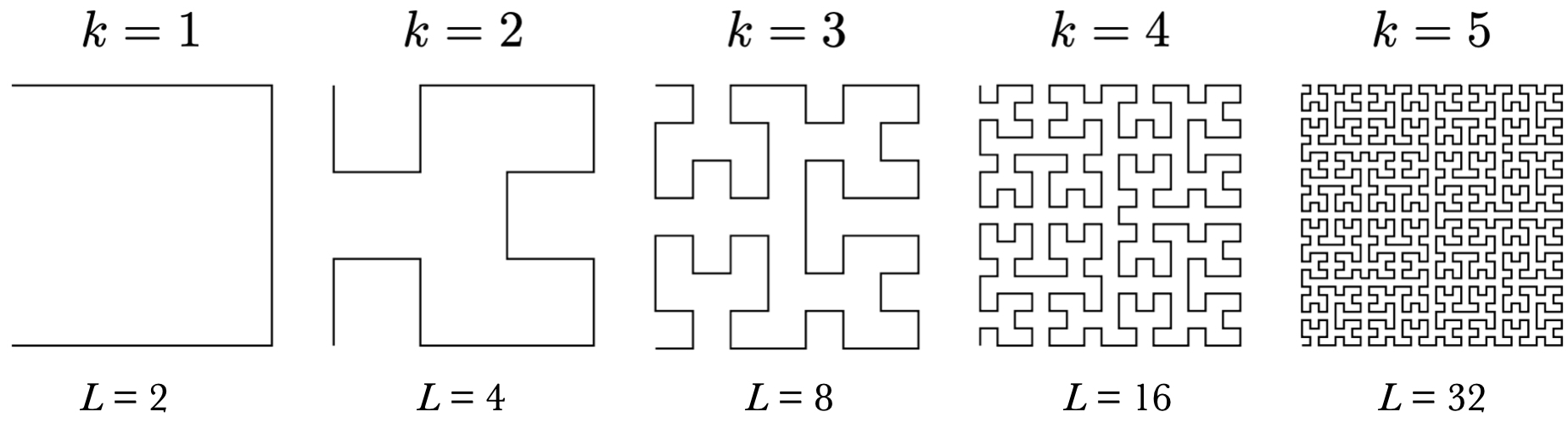} 
	\caption{Examples of Hilbert curves of order $k$ used for path encoding}
		\label{fig_hilbert_curve}
\end{figure}

Usually, a row by row flattening of the matrix is used, but more complex paths such as Hilbert, Peano or Moore curves \cite{sagan2012space} can be used. In \efig{fig_SARW}, for illustration purpose \footnotemark\footnotetext{It illustrates the fact that any space-filling curve could be in principle be used}, a fully-packed self-avoiding walk is generated by a pivot algorithm \cite{oberdorf2006secondary}.
Intuitively, the path should be as compact as possible such that local correlations in the image remain local upon transformation. In the following, we will use mostly Hilbert curves (see  \efig{fig_hilbert_curve}) unless specified otherwise.
This encoding allows for efficient quantum image processing by parallel manipulation of all pixel values through unitary operations on the quantum state. This way, we only need $2n+1$ qubits to store the information of $2^{2n}$ grayscale pixel values, which could lead (in the best cases) to an exponential reduction to the classical case. Although this encoding requires one more qubit that other encodings, it has been shown to work the best on natural images \cite{jobst2024efficient}. \textcolor{black}{The improved representation performance of FRQI over simple amplitude encoding comes from the factorization between spatial degrees of freedom and pixel values inherent to the FRQI construction.}

\begin{figure}[ht!]
\includegraphics[scale=0.25]{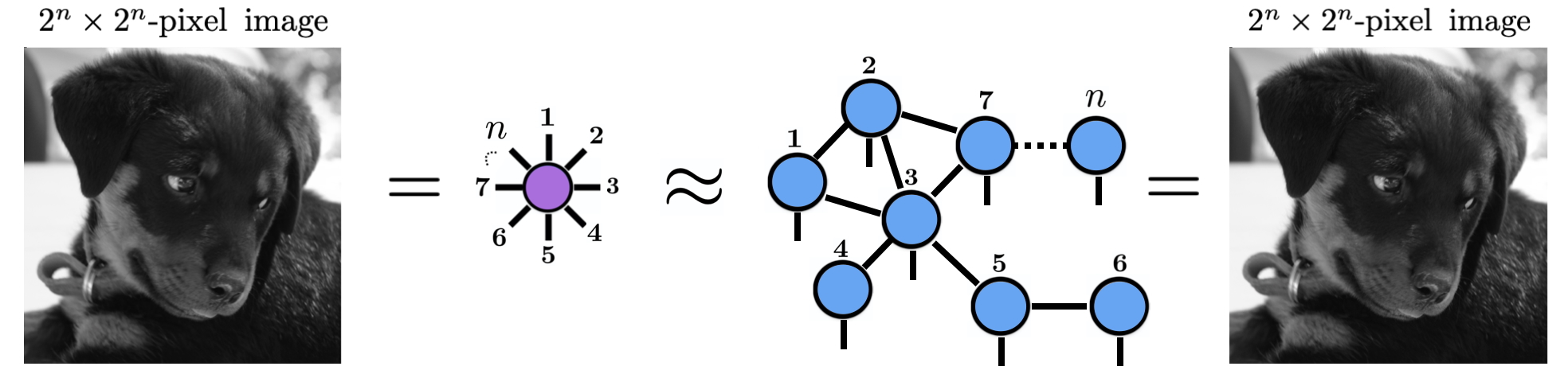} 
	\caption{Tensor network compression of a normalized black and white image $h_{ij}$. Here, a generic tensor network is shown for the sake of generality. The approximate symbol will quantify the precision of the compression. The final contracted object is the decompressed image $\tilde h_{ij}\approx h_{ij}$.}
		\label{fig_6}
\end{figure}

Thanks to the multi-scale nature of typical images, an other interesting and relevant way of encoding images into tensors is the so-called quantics representation \cite{oseledets2009approximation,khoromskij2011d} (first appeared in the quantum computing community \cite{wiesner1996simulations,zalka1998simulating,latorre2005image}) for the tensors as it naturally encodes scale separation. As a result, coarse features are encoded in high-order qubits, fine details live in low-order qubits and scale separation is built into the Hilbert space. It represents a real-space block renormalization expressed in terms of binary degrees of freedom
rather than continuous spatial coordinates.
\begin{figure}[ht!]
\includegraphics[scale=0.20]{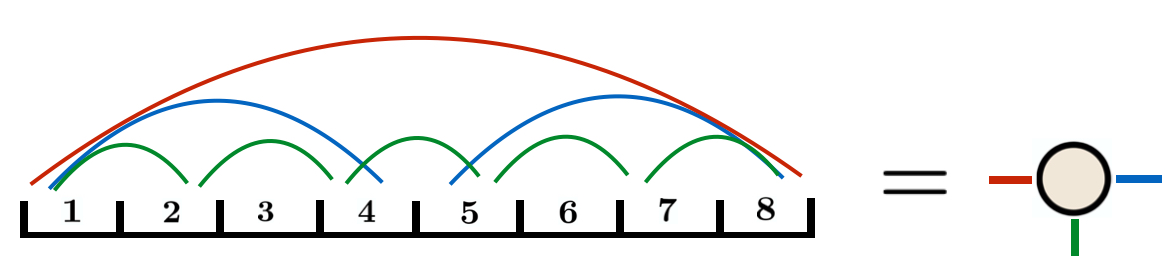} 
	\caption{Quantics representation of the vector $\vec{x}$ on $L=8$. Let us denote $\psi_{ijk}$ the resulting tensor components. In this example, we have $x_1\rightarrow \psi_{000}$, $x_2\rightarrow \psi_{100}$...,$x_7\rightarrow \psi_{111}$ and $x_8\rightarrow \psi_{001}$.}
\label{fig_5}
\end{figure}
Contrary to FRQI, the quantics representation does not flatten the matrix, but factorizes the indices. In $2d$, this means splitting the indices corresponding to spatial coordinates $x$ and $y$ in their binary components $(x_1x_2x_3...x_n)_2$ and $(y_1y_2y_3...y_n)_2$ and rearranging them to form new tensor indices $\mu_p=(x_p y_p)_2$. This yields a $2n$-order tensor $\psi_{\mu_1\mu_2...\mu_n}$ where each $\mu_p$ corresponds to a particular length scale. \footnotemark\footnotetext{As an example the value of the image at coordinate $(x,y)=(43,78)$ on a grid of size $L=128$ is given by the component $\psi_{00101011001110}$ of a 14-qubits state $|\psi\rangle$.} Technically a full tensor of the form $\psi_{\mu_1\nu_1\mu_2\nu_2...\mu_n\nu_n}$ contains $2^n\times 2^n$ parameters, as many as the initial matrix we are considering, only in a different tensor shape. 
\textcolor{black}{In both cases (quantics or qubit encoding), we end up with a full tensor that contains exactly the same information as the original image.} A non-trivial decomposition of this tensor into an interconnected network might reduce the number of parameters from an exponentially to (in the best cases) a logarithmically small number. This generic approach applied to our original image is illustrated in \efig{fig_6}. 
This tensor network decomposition could be used as a way of compressing images ($h_{ij}\rightarrow\tilde h_{ij}$) for transmission and storage or, as we will see later, as a crucial step in a pipeline of transformations allowing to outperform standard optical simulations or image processing. \textcolor{black}{The elephant in the room is that both methods are SVD-based so they require an exponential number of operations to generate the MPS such it might be impractical for very large images $n>30$, the alternative will be to use tensor-cross interpolation. This issue will not be discussed in more detailed but will be one of the focus of attention in the second half of this paper.}  

Many different topologies of tensor networks \cite{bridgeman2017hand,biamonte2017tensor,orus2019tensor} have been proposed over the years and became very powerful theoretical and numerical tools that allow to capture complex phenomena regularly found in the fields of condensed matter, high-energy physics or quantum computation \cite{montangero2018introduction,collura2024tensor}. In the following section, we focus on the two most common topologies of tensor networks, namely matrix product states and tensor trees and evaluate their expressive and compressive capabilities for increasing sizes using FRQI and quantics encodings.

\section{Matrix Product States and Tensor Tree Networks}

The specific tensor network we are focusing on first, is called a matrix product state (MPS) in the physics literature or a tensor train (TT) in the mathematical community. 
\begin{figure}[ht!]
\includegraphics[scale=0.2]{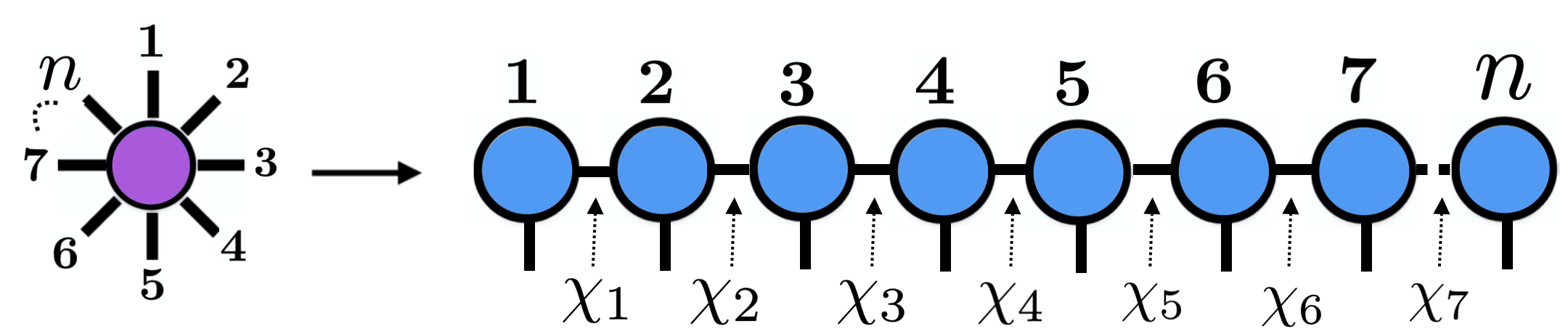} 
	\caption{Illustration of the decomposition of high-order tensor into a matrix product state. The arrow represents a sequential singular value decomposition and $\chi_k$ are the bond dimensions.}
		\label{MPS}
\end{figure}
\textcolor{black}{
The full $n$-qubit state vector $|\psi\rangle\in\mathcal{H}$, whether we choose the FRQI or quantics encodings \footnotemark\footnotetext{Technically the quantics representation does not lead to a proper quantum state while FRQI automatically does, although we can normalize the image before the encoding so we can use the notation $|\psi\rangle$ for both without confusion.}   is 
\begin{equation}
|\psi\rangle = \sum_{s_1, s_2, ..., s_n \in \mathbb{Z}_2} \psi_{s_1 s_2 ... s_n}|s_1 s_2 ... s_n\rangle,
\end{equation}
 where the tensor $\psi_{s_1 s_2 \dots s_n}$ is represented by the purple $n$-leg object in \efig{MPS}. The matrix decomposition outlined below is a generalization of the singular value decomposition and can be used to decompose a generic tensor into constituent tensors of lower order. Every bond corresponds to a Schmidt decomposition of the Hilbert space $\mathcal{H}=\mathcal{H}_{[1,2,...,k]}\otimes\mathcal{H}_{[k+1,k+2,...,n]}$. This succession ($\it{cf.}$ Appendix A for details) of operations leads to the usual form of the factorization of components
\begin{equation}
\psi_{s_1 s_2 \dots s_n} =
\sum_{\alpha_1, \dots, \alpha_{n-1}}
A^{[1]}_{s_1, \alpha_1}
A^{[2]}_{\alpha_1, s_2, \alpha_2}..
A^{[n]}_{\alpha_{n-1}, s_n},
\end{equation}
which is represented diagrammatically by the black object in \efig{MPS}. Depending on the scaling of $\chi_k$ with $n$, the MPS can efficiently represent $\psi_{s_1 s_2 \dots s_n}$. This point will be the focus on the next section.}

Once the image is encoded into the network, the numerical evaluation of the compressive power versus the reconstruction error is measured and the total number of parameters in the network is called the total dimension $D$. Now, we apply the decomposition to the picture showed earlier and analyze the error of the decomposition with respect to the original image for increasing allowed rank and increasing size. The goal is to mirror what we would have done with a typical stat-mech system, by studying the rank of the MPS for various sizes and infer the large-scale limit.

\textcolor{black}{Instead of using the quantum fidelity to evaluate the quality of the reconstruction, we define the mean squared error (as it is more common in machine learning) between the original matrix $h_{ij}$ and the decompressed image $\tilde h_{ij}$ as}
\begin{equation}
\epsilon(L)=\frac{1}{L^2}\sum_{ij}(h_{ij}-\tilde h_{ij})^2.
\end{equation}
\begin{figure}[ht!]
\includegraphics[scale=0.29]{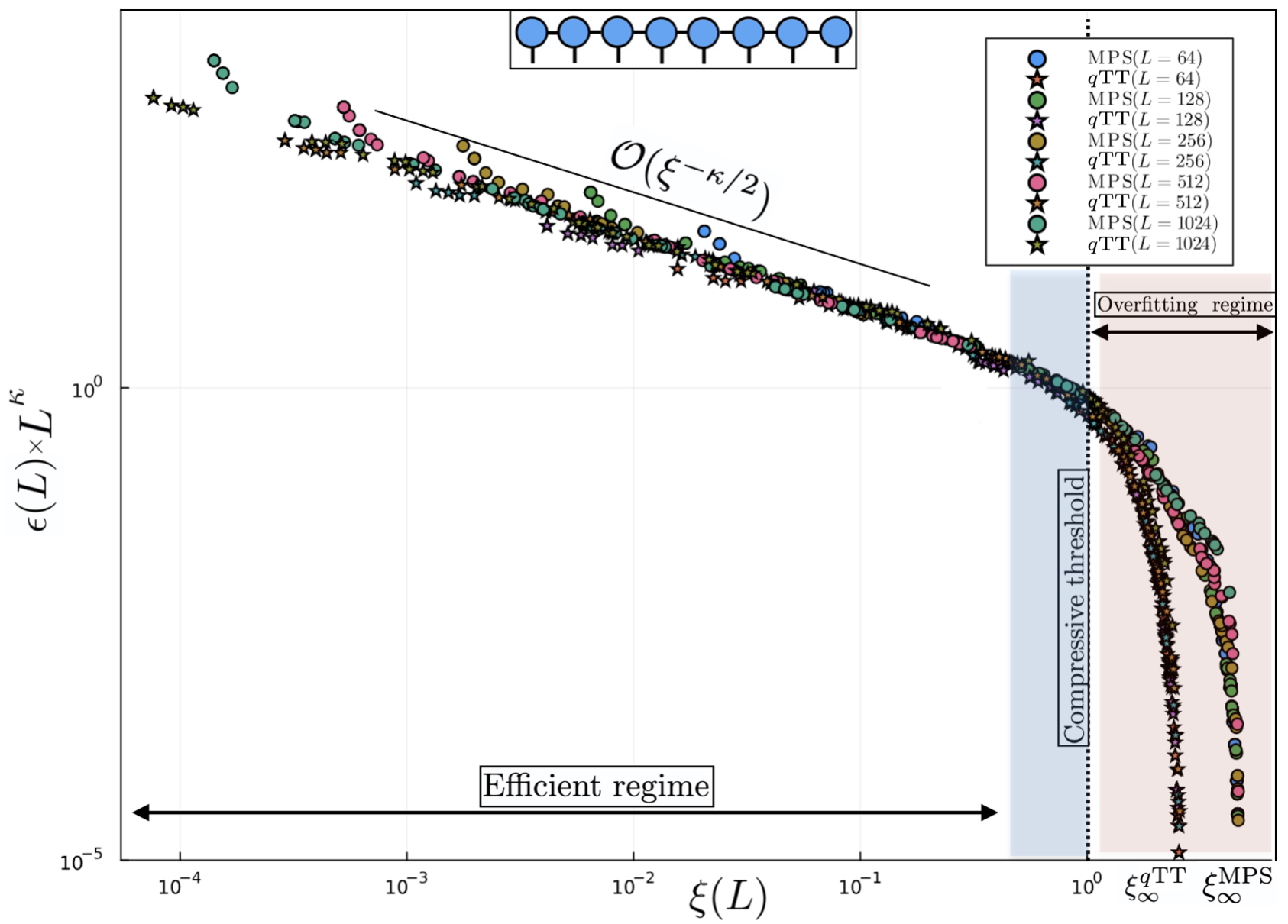} 
	\caption{Scaling of the mean squared error of the compression depending on system size for Hilbert-FRQI MPS and quantics tensor trains ($q\mathrm{TT}$).}
	\label{fig_scaling_mps}
		\label{fig_scaling_mps}
\end{figure}
Additionally, we introduce the inverse compression ratio as $\xi(L)=DL^{-2}$, which is the relevant order parameter that quantifies the efficiency of the approximation.
Numerical simulations (see \efig{fig_scaling_mps}) on qubit encoding matrix product states and quantics tensor trains suggest the following finite-size scaling
\begin{equation}\label{scaling}
\epsilon(L)=L^{-\kappa}\phi(\xi).
\end{equation}
with $\phi(\xi)\sim\xi^{-\alpha}$ with $\alpha\approx 0.75$ and $\kappa\approx 1.5$ for $\xi \ll1$. In this efficient regime $\epsilon(L)\sim\mathcal{O}(1)$, meaning that the error is (modulo multiplicative logarithmic corrections) independent of system size (consistent with \cite{jobst2024efficient}), and results are fairly similar between the two encodings. Conversely, the compression ratio at fixed error increases polynomially $\epsilon=C^{0.75}\times L^{1.5}$, so $C(L)\sim L^2$. This is in sharp contrast with local compressors like JPEG, since they possess a maximum compression ratio which is given by the division of the original image in small finite-size squares, this point will be made clearer later on.

\begin{figure}[ht!]
\includegraphics[scale=0.35]{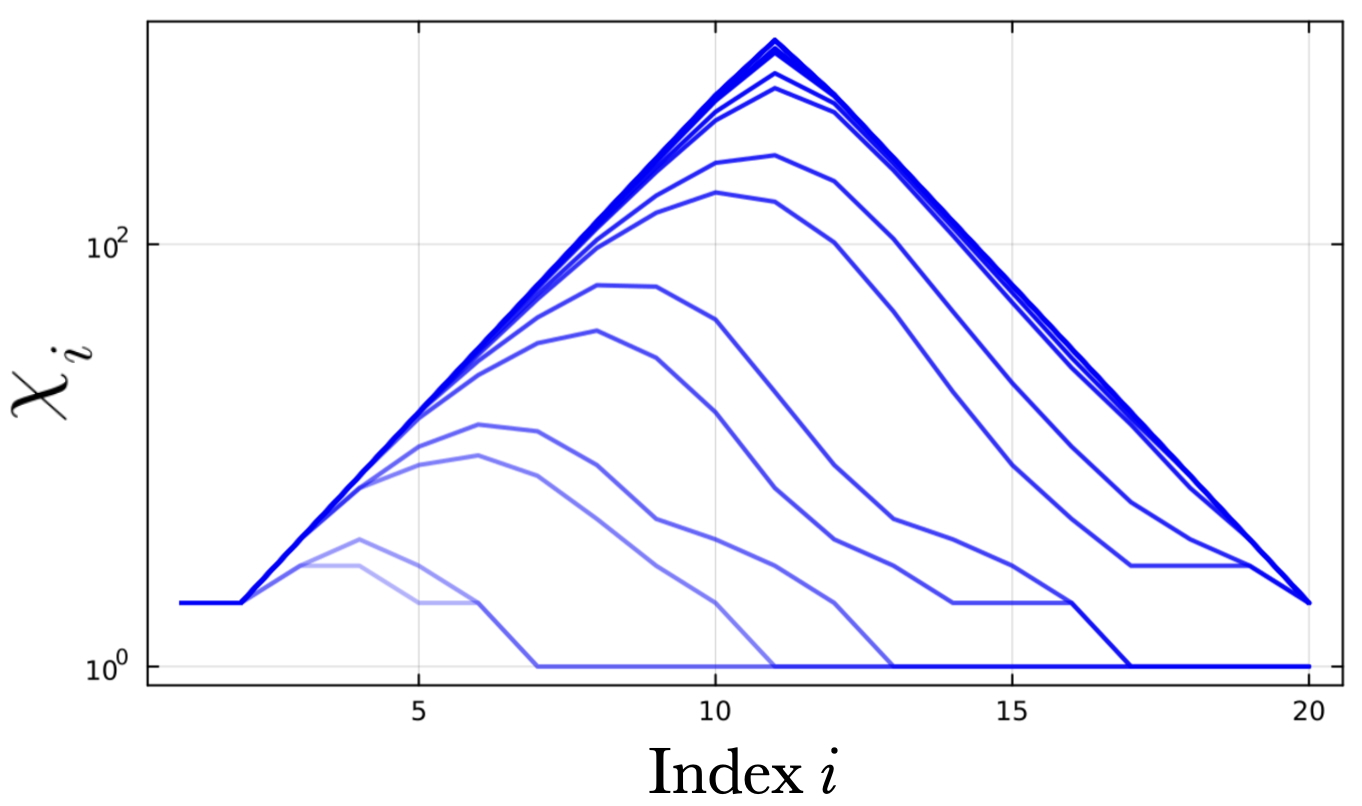} 
	\caption{Evolution of the bond dimensions along the $q\mathrm{TT}$ for $n=20$ for $\epsilon=10^{-1},10^{-2},...,10^{-12}$ (bottom to top curves). Scale separation is very noticeable at large error, while the ranks are maximized (Haar-like) for lossless reconstruction.}
	\label{pyramid}
\end{figure}

\textcolor{black}{We define $\xi(L)=1$ as the compressive threshold, as for $\xi(L)<1$ the approximation is compressive, while for $\xi(L)>1$ the number of parameters becomes greater than the original image.} In the overfitting regime (the red region $\xi(L)\geq1$), the error goes to zero exponentially quickly at a value $\xi\rightarrow\xi_{\infty}$. We notice that $\xi_{\infty}^{q\mathrm{TT}}<\xi_{\infty}^{\mathrm{MPS}}$ by a big margin (because there is one less qubit needed in the quantics representation than FRQI). This is due to the fact that, by construction of the singular value decomposition,  the dimension of a random MPS of $n$ sites can be showed recursively to be $D_\mathrm{max}(n)=8/3\times 2^n$. This can be found by integrating the exponential Page's curve \cite{page1993average}  of a Haar state over every bonds \efig{pyramid}.  It implies  $\xi_{\infty}^{q\mathrm{TT}}=8/3$ and $\xi_{\infty}^{\mathrm{MPS}}=16/3$ and at very high resolution the error is mostly dominated by pixel noise in the image. 

 \textcolor{black}{The linear network treats noise as global, since every bond is saturated.
Although interleaved quantics and FRQI encodings based on Hilbert curves differ
at the level of microscopic ordering, both induce a hierarchical dyadic
decomposition of the image and preserve locality across scales. As a result,
they lead to equivalent entanglement scaling and compression behavior when used
with linear tensor networks.} Interestingly, the value of $\xi_{\infty}$ is also of certain relevance in the intermediate regime (black region in \efig{fig_scaling_mps}) near the compressive threshold $\xi=1$ since a smaller value $\xi_{\infty}$ seems to force the curve to accelerate faster towards this line\footnotemark\footnotetext{A more fondamental question is whether or not a real image of the physical world (not a digital picture) would admit a true area-law at any reasonable scale.}. This fact allows the $q\mathrm{TT}$ to improve over the Hilbert MPS in this intermediate region. This argument will be made more clear in the following section.  

Another important decomposition is the binary tree tensor network (TTN) where the constituent tensors of the network lie on the nodes of a connected acyclic tree graph. The appeal of tensor tree network comes down to their relative ease of constructions and contractions as well as their greater expressive power and has been recently explored in the context of multivariate function approximations \cite{tindall2024compressing}.
\begin{figure}[ht!]
\includegraphics[scale=0.25]{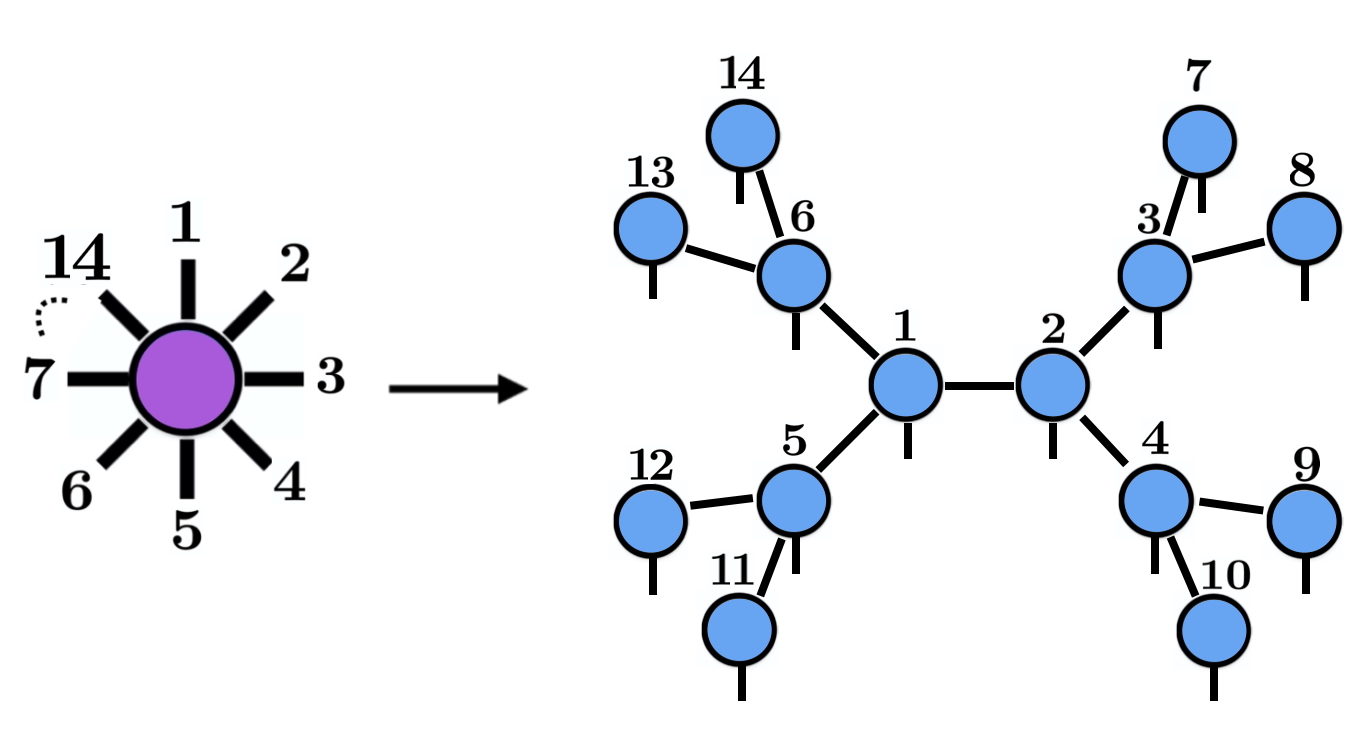} 
	\caption{Left: Binary tensor tree network $\mathrm{TTN}_{B}$ with 14 physical bulk sites.}
	\label{fig_tree}
\end{figure}
\textcolor{black}{The construction of tree tensor networks is not excessively different than the case of matrix product states. Once fixed a binary tree topology, every SVD corresponds to a tree cut which can separate non-contiguous qubits in the original ordering $\{1,2,...,k\}|\{k+1,k+2,...,n\}$. We can decompose the full tensor into a binary tree of lower-order tensors using a sequence of Schmidt decompositions corresponding to the cuts defined by the edges of the tree. Every edge $e$ of the tree defines a bipartition of the Hilbert space
\begin{equation}
\mathcal{H} = \mathcal{H}_{\text{subtree}(e)} \otimes \mathcal{H}_{\text{rest}(e)}.
\end{equation}
The construction is detailed in appendix A and a factorization of each subtree recursively into tensors associated with each node of the binary tree produces the usual tree decomposition of the tensor components
\begin{equation}
\psi_{s_1 s_2 \dots s_n} =
\sum_{\{\alpha_e\}} \prod_{v \in \text{nodes}} A^{(v)}_{s_v, \alpha_{v_1}, \alpha_{v_2}, \alpha_{v_\text{parent}}},
\end{equation}
where each $A^{(v)}$ has indices for the physical qubit at node $v$, the virtual bonds to its children, and the bond to its parent (if any). This is the direct generalization of the MPS canonical form to a binary tree geometry.}

\begin{figure}[ht!]
\includegraphics[scale=0.275]{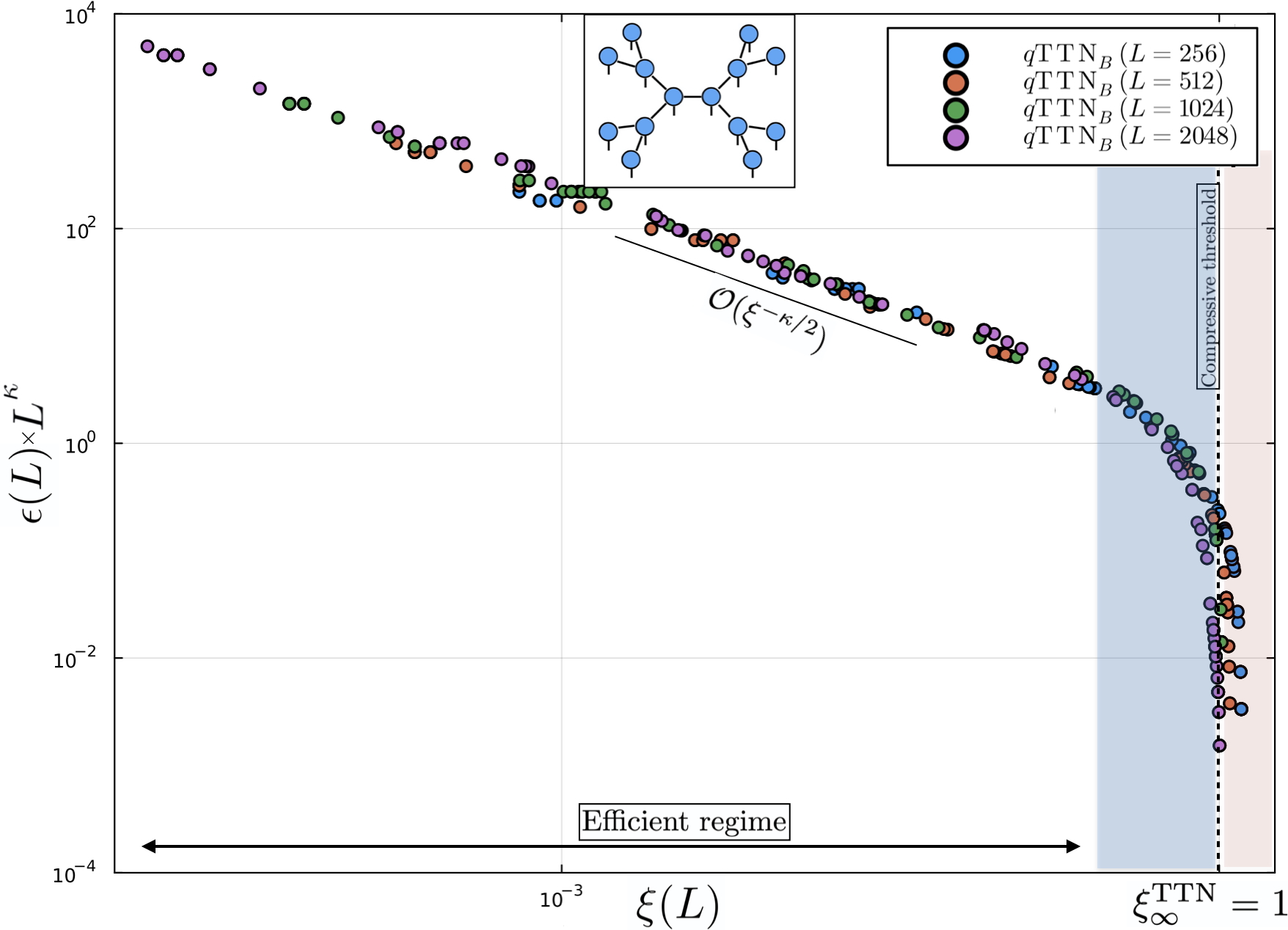} 
	\caption{Scaling of the error with the system size for $q\mathrm{TTN}_{B}$.}
	\label{fig_scaling_ttn}
\end{figure}

For notation convenience, we introduce the notation $q\mathrm{TTN}_{B}$ for quantics encoding on this tree. In \efig{fig_scaling_ttn}, we plot the scaling of $\xi(L)$ for $q\mathrm{TTN}_{B}$ and observe a similar scaling as before. \textcolor{black}{In quantum many-body physics, it is common knowledge that matrix product states can efficiently represent $1d$ quantum systems with area-law entanglement. In contrast, the hierarchical structure of tensor trees can also describe more complex types of entanglement, including some long-range quantum correlations, in higher-dimensional or critical systems where entanglement scaling can exceed that of an area law. In the scenario that we are interested in, the image considered follows an area law (since the bond dimensions of the MPS remain constant up to quite a large precision), and tensor trees only marginally improve the compression ratio but not the scaling of the function $\xi(L)$.}

Nevertheless, it is clear that the error goes to zero rather differently due to the maximal number of parameters allowed by the SVD in these kinds of networks. We can show that a large $n$, a maximally random tensor tree has at most $2^n$ parameters. We take the example of the boundary binary tree (see right \efig{fig_tree}), the maximal number of parameters is $D_\mathrm{max}(n)= n(2\times2) + \sum_{k=1}^{\log_2(n) - 2} \frac{n}{2^k}\left( 2^{2^{k+1} - 2}\times2^{2^{k+1} - 1} \right)+2\times2^{n-2}$. The first term $n(2\times 2)$ corresponds to the $n$ outside branch tensors and the last term corresponds to the two central tensors. The largest term inside the sum is $2^n-1$ so $D_\mathrm{max}(n)=4n+2^{n-1}+2.2^{n-2}=2^n+4n$, implying a value (up to a tiny $\mathcal{O}(n)$ term) of $\xi_{\infty}^{\mathrm{TTN}}=1$ for large $n$. \textcolor{black}{The exponential term 
comes from top layers, required to encode maximal entanglement and 
the linear term $4n$ comes from leaves and small subtrees. The noise is treated as scale-separated, since it lives un in the tree.}

 A similar counting operation can be done for bulk tensor trees, and would have lead to the same result. \textcolor{black}{It implies that in the thermodynamic limit (in contrast to matrix product states), tensor trees always compress the image, $\xi^{\mathrm{TTN}}_\infty\leq1$ regardless of the level of approximation (and encodings) which is already an interesting fact on its own and will be further explain in the following}. Therefore, tensor trees offer greater compressive power in the intermediate regime (see \efig{fig_scaling_ttn}), as the number of parameters approaches the compression limit with increasing system size. This is illustrated in the table below, where tensor trees clearly outperform linear networks and continue to compress the image even for very small reconstruction error\footnotemark\footnotetext{We do not show the results of FRQI on trees as they perform poorly compared to quantics.}. 
 \begin{table}[ht!]
\centering
\begin{tabular}{ ||c | c|c|c| c|| } 
\hline
 Tensor Nets and encodings& $\epsilon=10^{-3}$ & $\epsilon=10^{-4}$ & $\epsilon=10^{-5}$ \\
\hline
MPS (SARW) & $270$ & $1.46$ &$0.48$  \\
MPS (S) & $320$ & $2.24$ &$0.46$  \\
MPS (H) & $330$ & $4.1$ &$0.64$   \\
$q\mathrm{TT}$ & $352$ & $4.2$ &$0.72$  \\
$q\mathrm{TTN}_{B}$  &$\bold{370}$ & $\bold{10.2}$ &$\bold{2.3}$  \\
\hline
\end{tabular}
\caption{Compression ratio for several errors with $L=2048$. The three firsts rows use FRQI with respectively a random walk (SARW), snake-like (S) and a Hilbert curve (H) path encoding. The fourth is the quantics tensor train $q\mathrm{TT}$ and the last is a quantics encoding on bulk tensor trees $q\mathrm{TTN}_{B}$.}
\label{table_results}
\end{table}

\textcolor{black}{In the case of trees, entanglement is distributed across a hierarchy of scales rather than along a single ordered chain, and is progressively pushed toward higher levels of the tree as reconstruction accuracy increases. When encoding an image using the quantics indexing, the resulting tensor indices are naturally ordered by spatial scale, from coarse to fine resolution. While this multi-scale indexing already improves the entanglement structure compared to a path encodings, the topology of the tensor network remains the main factor for compression. In a MPS, the chain topology forces noise-induced entanglement to be distributed uniformly across all links, leading to large bond dimensions at all scales.}

\textcolor{black}{By contrast, a binary TTN aligns naturally with the hierarchical structure induced by the quantics representation: local tensors capture fine-scale correlations, while global correlations and noise entropy are progressively pushed toward the upper layers of the tree. As a result, Haar entanglement appears only at the highest levels of the TTN, allowing strong and aggressive truncation of top bonds while preserving local image features. This scale separation of entropy explains why TTN representations provide more efficient compression of noisy images than MPS, even when both use the same quantics encoding.
Since TTN is strictly more efficient in the noisy regime while remaining
equivalent to MPS in the area-law limit, continuity of entanglement scaling
implies that TTN is generically advantageous in the intermediate regime where
weak violations of the area law are present, as is typical for noisy or
high-precision data. When the entanglement structure becomes genuinely
2$d$ or exhibits scale-invariant behavior, PEPS and
MERA offer more natural and efficient descriptions by adapting the tensor
network geometry with the underlying entropy structure of the data.}

\textcolor{black}{Although comparisons with standard image compression algorithms are not showed here, we mention that tensor networks might have a big advantage at very large size and relative high error. Classical block-based compression schemes such as JPEG operate by partitioning
the image into fixed-size blocks and applying local transforms within each
block, thereby imposing a hard cutoff on correlations beyond the block scale.
While effective for locally smooth images, this rigid locality prevents an
adaptive representation of long-range or multi-scale correlations and leads to
well-known blocking artifacts. In contrast, tensor network representations
do not enforce fixed spatial partitions but instead truncate correlations
adaptively across scales, allowing global structure to be preserved while
suppressing noise.  A comprehensive benchmark, although beyond the scope of this paper, on the performance of tensor networks across large datasets of diverse images would be necessary to rigorously demonstrate their relative strengths and weaknesses.}

Despite these promising features, one of the main strength of tensor networks lies not only in data reduction itself but in their ability to perform complex operations efficiently. In particular, there are particularly suited for image transformations tasks, where operations can be applied directly at the level of the tensor network. The last section of this work will therefore be directed towards optical applications that could be significantly sped up using this formalism.

\section{Image processing and optical image formation}

\begin{figure}[ht!]
\includegraphics[scale=0.16]{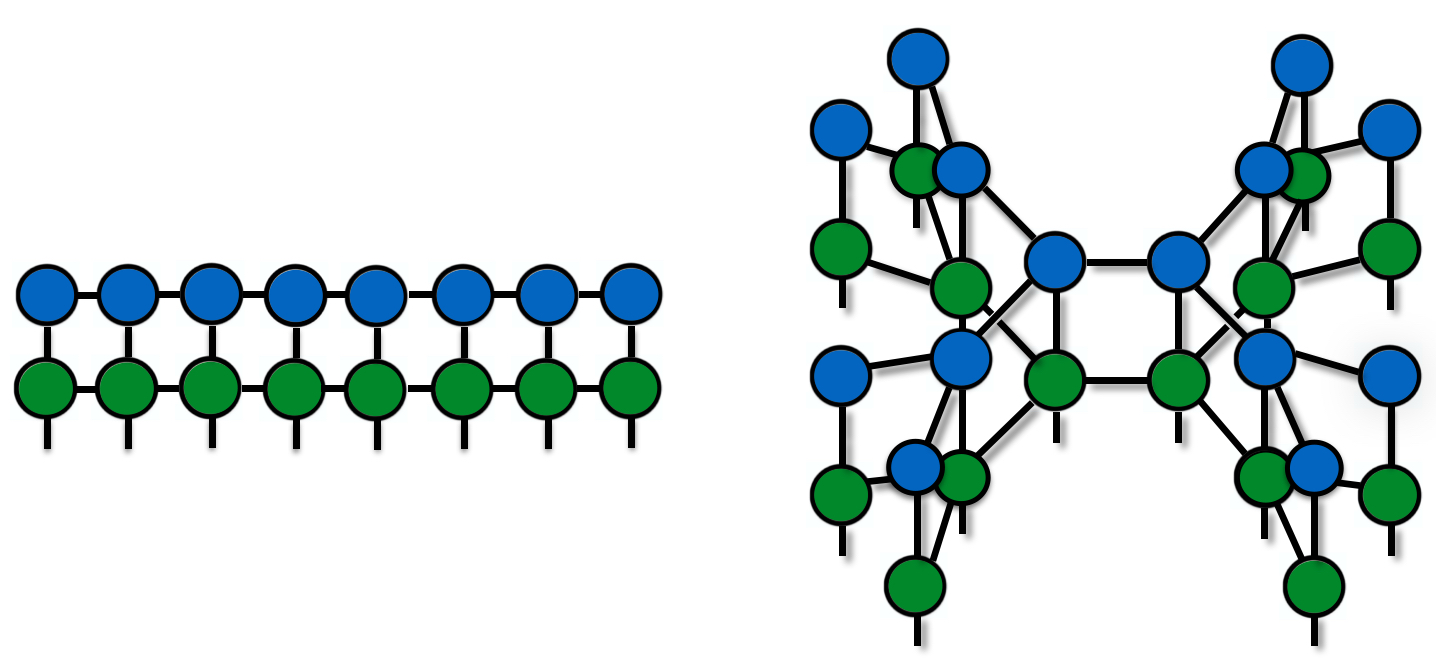} 
	\caption{Operations on images as MPS/MPO contraction on the left and TTN/TTO contraction on the right.}
	\label{fig_mpo}
\end{figure}

Many geometric operations on images, such as shifts, rotations and scale transformations can be represented efficiently by tensor network operators of small and fixed rank. In the case of MPS (TTN), the operators are appropriately named matrix product operators (tree tensor operators). Those objects are tensor networks that can act on other tensor networks (see \efig{fig_mpo}).  The complexity of the contraction MPS/MPO (TTN/TTO) depends strongly of the rank of the two objects $\mathcal{O}(2n\chi^2_1\chi^2_2).$ This favorable scaling allows for very efficient operations that would have been very costly otherwise.
Similarly, image transforms such as the Hadamard transform, discrete Fourier transform or the discrete sine/cosine transforms can also be constructed very efficiently in this formalism. Quantum mechanically speaking, it means that the equivalent quantum circuit does not create a lot of entanglement among the qubits, a weakly entangled initial state $|\phi\rangle$ propagating into the circuit would result in another (perhaps even more) weakly entangled state $|\psi\rangle$. 

A trivial (but still useful) example of operator with specific applications in image processing and compression is the previously mentioned Hadamard transform.
The Hadamard matrix of order $ 2^n $, denoted $ H_n $, can be defined recursively using the tensor product 
\begin{equation}
H_1 = \frac{1}{\sqrt{2}} \begin{pmatrix}
1 & 1 \\
1 & -1
\end{pmatrix}, \qquad
H_n = H_1 \otimes H_{n-1} \quad \text{for } n \geq 2.
\end{equation}
Equivalently, it can be written as a unitary operator $\hat{U}$ on the space of qubits such that\footnotemark\footnotetext{In this paper, we choose the quantum computation notation $\{\hat{X},\hat{Y},\hat{Z}\}$ instead of $\{\hat{\sigma}_x,\hat{\sigma}_y,\hat{\sigma}_z\}$ for the Pauli matrices.} 
\begin{equation}\label{hadamard}
\hat{U} = H_1^{\otimes n}=\left(\frac{\hat{X}+\hat{Z}}{\sqrt{2}}\right)^{\otimes n},
\end{equation}
which is by definition a matrix product operator of rank $1$ in terms of Pauli gates. Classically, the Hadamard transform can be computed in ${\displaystyle L\log L}$ operations, using the fast Hadamard transform algorithm. With tensor networks (or a quantum computer, as it is a quantum logic gate that can be parallelized) it would take $\mathcal{O}(1)$ time.
Because of the trivial structure, the Hamiltonian $\mathcal{H}$ and the evolution operator ($\mathcal{H}\propto\mathbb{I}-\hat{U}$) are the same up to a phase.  Explicitly, the MPO governing the unitary evolution reads
\begin{equation}\nonumber
\hat{U}= \sum_{\substack{a_1, \ldots, a_n = \{0,1\} \\ b_1, \ldots, b_n = \{0,1\}}} 
\left( \prod_{i=1}^n W^{[i]}_{a_i b_i} \right)
\left|\mathbf{a}\right\rangle \left\langle \mathbf{b}\right|,
\end{equation}
where the local tensors
\begin{equation}
W^{[i]}_{a_i b_i} = H_1\delta_{\alpha_{i-1},1}\delta_{\alpha_i,1},
\end{equation}
are of order $(2,2,1,1)$.  The same local tensors can be similarly plugged on a tree tensor network such as shown in \efig{fig_mpo}. Applying the Hadamard transform to an image approximated by an MPS or TTN is simply the application of a Hadamard gate to each qubit individually. This suggests that some image processing tasks or optical calculations that can be expressed as tensor operators might be significantly faster than more standard classical methods. Complex Hadamard matrices in general cannot be written as \eref{hadamard}, and a description in terms of MPO would surely lead to an exponential increase in bond dimensions. As a counter point, although the discrete Fourier transform is a particular case of complex Hadamard matrices, it has been shown that the quantum Fourier transform operator can be written explicitly as a matrix product operator with a finite rank $\chi$ and error $\epsilon=\mathcal{O}(n \mathrm{e}^{-\chi\mathrm{log}(\chi/3)})$ \cite{chen2023quantum,chen2024direct}. The last section of this article will be devoted to optical applications of such speed-ups. 

\begin{figure}[ht!]
\includegraphics[scale=0.075]{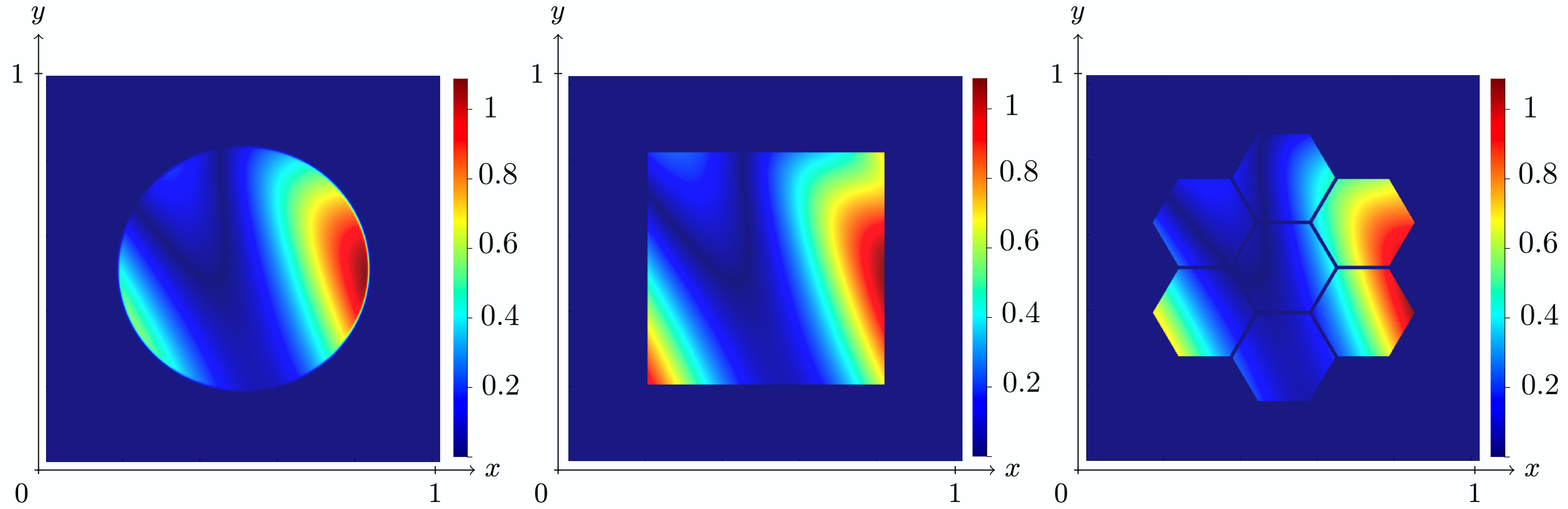} 
	\caption{Three classic apertures considered for the simulations modulated by continuous phase field living on them. The middle case will be surprisingly easy to approximate with quantics TCI while the others will turn up to be more difficult.}
	\label{fig_aperture}
\end{figure}

In the following, we shall see how to incorporate our previous results on image encoding in the framework of a realistic optical system. The simplest of calculations is to compute the point spread function (PSF) of a classical optical system that contains phase aberrations. An image observed with that aberrated optical system will then be modeled by the convolution of the input image and the PSF. Let us start be defining a monochromatic coherent field as the plane wave
\begin{equation}\label{field}
E_0(r, \theta) = A(r, \theta) \mathrm{e}^{-ik \Phi(r, \theta)},
\end{equation}
where $k=2\pi/\lambda$, $A(r, \theta)$ is the amplitude of the wavefront describing the intensity variations and
$\Phi(r, \theta)$ is the phase function, which captures the optical path difference and aberrations of the optical system.  
Phase aberration $\Phi(r, \theta)$ is often expanded using Zernike polynomials $\Phi(r, \theta) = \sum_{n,m} c_n^m Z_n^m(r, \theta)$,
where $ Z_n^m(r, \theta) $ are the Zernike polynomials, and $ c_n^m $ are the corresponding coefficients representing different aberrations like defocus, astigmatism, and coma. The focal plane amplitude distribution is obtained from the Fraunhofer diffraction integral (see the standard textbook \cite{goodman2005introduction})
\begin{equation}
E_{\mathrm{f}}(u, v)=\int r \mathrm{d}r\int\mathrm{d}\theta E_0(r, \theta) \mathrm{e}^{-i k r (u \cos\theta + v \sin\theta)},
\end{equation}
where $(u, v)$ are spatial coordinates in the focal plane, and $A(r, \theta)$ is the pupil function, typically a circular aperture that \footnotemark\footnotetext{We switch liberally from Cartesian to polar coordinates depending on cases.}.  
\textcolor{black}{For simplicity in the simulations, we choose the arbitrary function $\Phi(u,v) = \cos(u)\cos(u) + \cos(2u + 5v) + \sin(u v) + \cos(2u + 2v)^{13}$ and the apertures showed in  \efig{fig_aperture}}. The point spread function is simply given by 
\begin{align}
\text{PSF}(u, v) &= \left| E_{\mathrm{f}} (u, v) \right|^2\nonumber\\
&=|\mathcal{F}(E_0)|^2,
\end{align}
which is the intensity of the field at the focal plane ($\mathcal{F}$ being the Fourier transform). In the following we will consider it as normalized $\text{PSF}(u, v)\rightarrow \text{PSF}(u, v)/|\text{PSF}(u, v)|$ such that it represents a photon probability distribution.

\begin{figure}[ht!]
\includegraphics[scale=0.25]{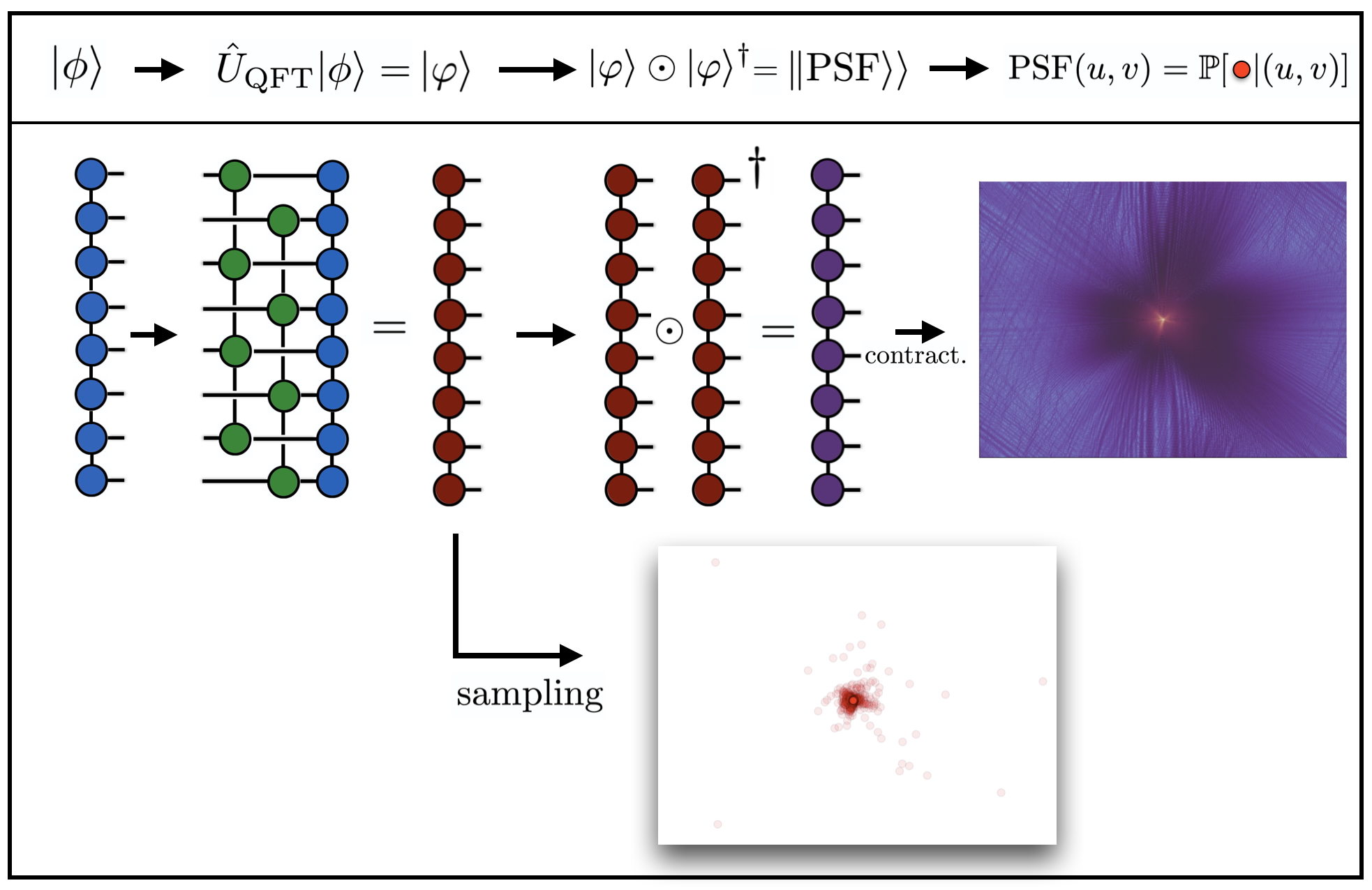} 
	\caption{Schematic representation of the tensor construction of the PSF computed with the circular aperture showed in \efig{fig_aperture}. The Fourier MPO is applied twice, first in the $x$ then $y$ direction.}
	\label{fig_psf_mps}
\end{figure}

\textcolor{black}{Let us see how one defines this classical calculation in terms of MPS and MPO. For the time being, we admit that we successfully constructed a normalized MPS representation of $E_0$ that we note $|\phi\rangle$. The classical Fraunhofer propagation of the continuous field $E_0$ can be seen as a unitary propagation of a normalized quantum state $|\phi\rangle$ living in the Hilbert space $\mathcal{H}$ of infinitely many qubits. The non-unitary operation that computes the PSF of the optical system is, in this context, defined by the non-unitary process of quantum measurement.  Mathematically, it boils down to the following dictionary
\begin{align}\label{psf_eq}
&E_0 \leftrightarrow |\phi\rangle\in\mathcal{H}, \nonumber\\
&\mathcal{F}_{xy}\{\bullet\}  \leftrightarrow \hat{U}_{\mathrm{qft}}=\hat{U}^{(x)}_{\mathrm{qft}}\otimes\hat{U}^{(y)}_{\mathrm{qft}}:\mathcal{H}\rightarrow\mathcal{H},\nonumber\\
&E_{\mathrm{f}} \leftrightarrow |\varphi\rangle=\hat{U}_{\mathrm{qft}}|\phi\rangle\in\mathcal{H},\nonumber\\
& \mathrm{PSF} \leftrightarrow \|\mathrm{PSF}\rangle\rangle
= |\varphi\rangle \odot |\varphi\rangle^{\dag}\in\mathcal{H}_{\mathrm{cl}},
\end{align}
where we use $\|.\rangle\rangle$ to denote MPS representations of classical fields or probability distributions living on $\mathcal{H}_{\mathrm{cl}}$, as opposed to quantum state vectors in $\mathcal{H}$ and where $\odot$ is the local contraction of physical indices, corresponding in the continuum limit to point-wise multiplication in the measurement basis \footnotemark\footnotetext{The algorithm used to perform the calculation of element-wise multiplication scales as $\mathcal{O}{(d\chi^3)}$, a faster algorithm \cite{michailidis2024tensor} could be implemented to speed things up.}.  The local contraction \footnotemark\footnotetext{In this notation, $|\varphi\rangle \odot |\varphi\rangle^{\dag}$ stores the diagonal values of the density matrix $\rho$ such that $\mathcal{H}_{\mathrm{cl}} \simeq \{ \operatorname{diag}(\rho) : \rho \in \mathcal{B}(\mathcal{H}) \}.
$}  is the MPS analog of projective measurement, mapping a quantum state to a classical probability $\mathrm{PSF}=|\varphi|^2$, with normalization $\langle\varphi|\varphi\rangle = 1$ (equivalently $\|\mathrm{PSF}\|_{L^1}=1$).
This normalized PSF then can be seen a the spatial probability of photon events $\mathbb{P}[$\textcolor{red}{$\bullet$}$,(u,v)]$ where \textcolor{red}{$\bullet$} is the photon event.  A summary of the discretized calculation steps is illustrated in \efig{fig_psf_mps} \footnotemark\footnotetext{mention that a $\mathrm{FFTSHIFT}$ operation..)}}.

Let us mention, that the point-wise product operation is only needed if we absolutely require the PSF to be in a MPS format, to be applied later on to another MPO for example. But many operations can be done at the level of $|\varphi\rangle$. Direct contraction of the MPS $|\varphi\rangle$ followed by an absolute square operation can be done to plot quickly some parts of the large PSF for example. Also,  contrary to the standard FFT calculation, the matrix product state formulation allows photon sampling without ever computing the photon statistical distribution (see \efig{fig_psf_mps}) but by sampling the field $|\varphi\rangle$ after orthogonalization. The perfect sampling algorithm can be used to sample from this distribution \cite{stoudenmire2010minimally,ferris2012perfect}. As the name suggests, the algorithm draws perfect samples from photon distribution and is very efficient as it scales as $\mathcal{O}{(\chi^2)}$ for a properly right-orthogonalized MPS. This method could be especially useful in situations in which photon statistical behavior needs to be taken into account, or to construct a joint distribution over photon events and aberrations in a bayesian fashion.

 As said earlier, a low-rank matrix product operator construction of $\hat{U}_{\mathrm{qft}}$ has recently been developed and could be implemented here. To take advantage of this MPO formulation, the field $|\phi\rangle$ requires also a low-rank form.
In this section, we proceed to approximate the continuous function $E_0(r, \theta)$ by a MPS, without executing the exponentially expensive operation that is exhaustively storing the function on a grid $L\times L$. To overcome this limitation, we then adopt a sampling-based approach exploiting the family of tensor-cross interpolations algorithms. In particular, we focus on the quantics tensor cross-interpolation ($q$TCI) method described in \cite{ritter2024quantics,nunez2025learning}, which is based on the LU decomposition. The algorithm we employ takes the input, represented as a function, and generates an approximated MPS in the quantics interleaved representation. The final tensor is obtained by deterministically sampling the initial one, with a linear number of calls $\mathcal{O}(n\chi^2)$, depending on the rank of the function.

We now proceed to approximate the amplitude $A(r,\theta)$ with a quantics tensor train. For our simulations, we chose three relevant geometries   a very favorable, a very unfavorable and a fairly middle of the road (see \efig{fig_aperture}) in order to identify the pros and cons of these methods. The problem of discretizing a circular discontinuous function is already discussed in Fernandez et al \cite{nunez2025learning}, they showed that the bond dimension increases along the tensor train as $\chi_p\sim 2^{n/2}$ with a maximum bond dimension of $\chi_\mathrm{max}\sim2^{2(n+1)/3}$. This behavior is correct for the exact discontinuous function but the bond dimension can be reduced by using a smoother function with a parameter dictating the smoothness of the discontinuity. For the following numerical simulations we choose a continuous function $A(r,\theta)=(1+\alpha^{-1}\exp((r-r_0)^2))^{-1}$ with $\alpha=0.001$ and $r_0=0.5$. In addition, we will also consider the two other geometries of \efig{fig_aperture}, namely a simple box function $A(r,\theta)=\mathrm{box}(x,y)$ and a segmented aperture  $A(r,\theta)=\mathrm{Hex}(x,y)$. 
 
\begin{figure}[ht!]
\includegraphics[scale=0.27]{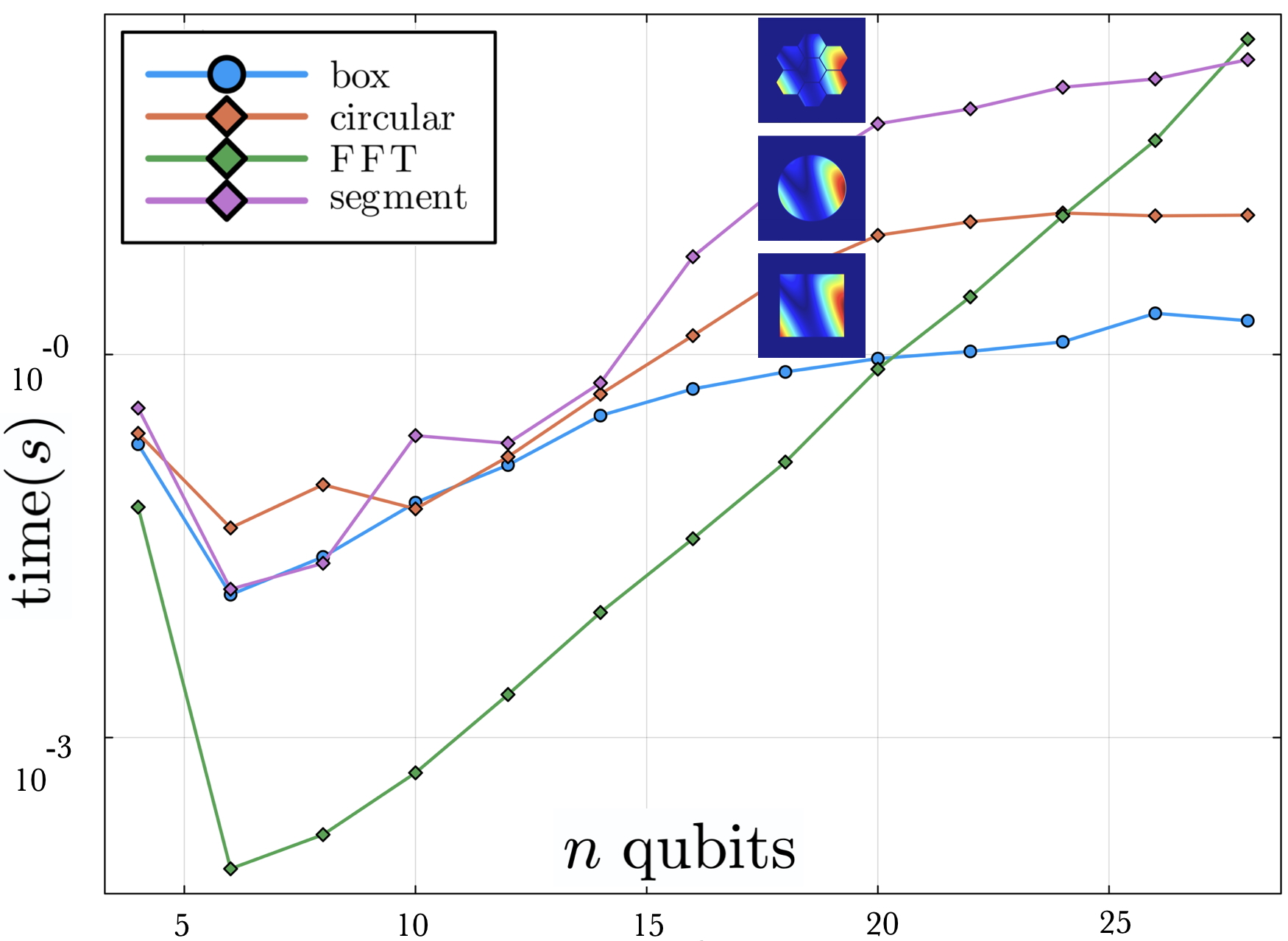} 
	\caption{Run times of \eref{psf_eq} for the three apertures showed in \efig{fig_aperture} compared to the fast Fourier transform (FFT) computation in green.}
	\label{fig_time_psf}
\end{figure}
It can be argued that since the phase function also live on the aperture, the maximum bond dimension is somehow dominated by the discontinuity, regardless of the degree of the polynomials considered. The numerical experiment (see \efig{fig_time_psf}) shows run times of \eref{psf_eq} for the three apertures shown earlier, including time necessary to encode the field \eref{field} with $q$TCI. Times are measured for several grid sizes defined by the number of qubits and are compared to a standard FFT calculation \footnotemark\footnotetext{\textcolor{black}{Package \texttt{FFTW} {https://github.com/JuliaMath/FFTW.jl}}}. In logarithmic time, it indicates that the run times for the MPS/MPO formulation are (up to log corrections) constant, with a constant depending on the complexity of $|\varphi\rangle$, while the FFT computation scales linearly with system size. \textcolor{black}{In the figure above, we stop at around $n=25$ but size upwards of $n>50$ are possible since we are working with $q$TCI and need not store an exponential number of values. This would not be the case in the next section since we will connect to the first section of this article and apply this formalism to the original image.}

\begin{figure}[ht!]
\includegraphics[scale=0.26]{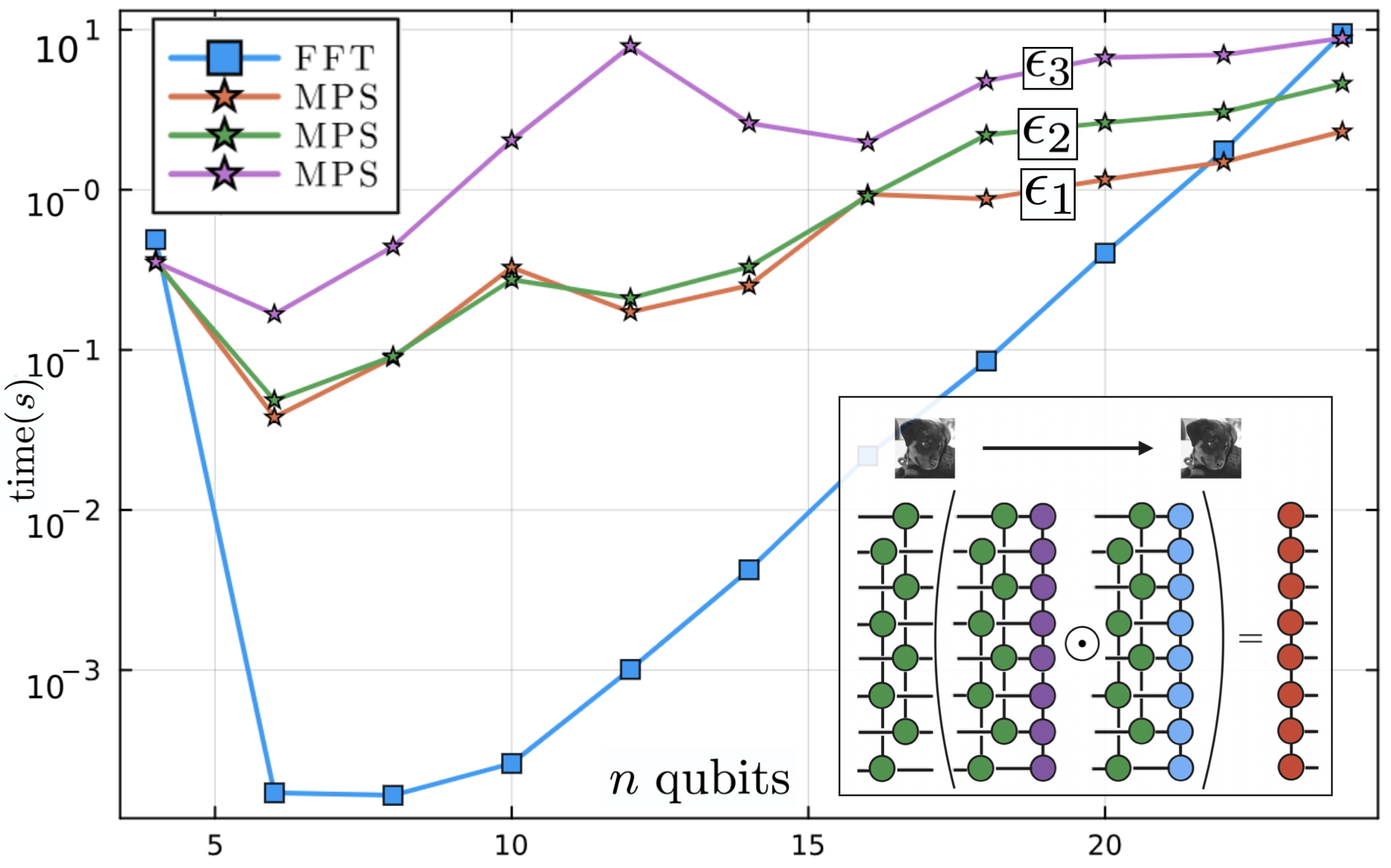} 
	\caption{Run times of the tensor network image formation for increasing precisions ($\epsilon_1=10^{-2}$,$\epsilon_2=10^{-3}$,$\epsilon_3=10^{-4}$) versus the standard FFT implementation. The MPS/MPO construction is shown on the bottom right and the circular aperture is used.}
	\label{fig_convolution}
\end{figure}

For an extended scene $ I_{\text{obj}}$, the image formed in the focal plane is given by the convolution of the scene with the point spread function. Mathematically, this is expressed as $I_{\text{img}}=I_{\text{obj}} \ast\text{PSF}$. In the Fourier domain, this convolution simplifies to a multiplication $\hat{I}_{\text{img}}(f_x, f_y) = \hat{I}_{\text{obj}}(f_x, f_y) \cdot \hat{\text{PSF}}(f_x, f_y)$,
where $ \hat{I} $ and $ \hat{\text{PSF}} $ are the Fourier transforms of the object and PSF, respectively. The function $\hat{\text{PSF}}$ is also known as the optical transfer function, which characterizes how different spatial frequencies are transmitted by the optical system. As we saw earlier, the extended scene $I_{\text{obj}}$ can also be seen as a normalized quantum state MPS $|S\rangle$ such that the final image state reads
\begin{eqnarray}\label{convolution_eq}
I_{\text{img}}=I_{\text{obj}} \ast\text{PSF}\rightarrow\hat{U}_{\mathrm{iqft}} (\hat{U}_{\mathrm{qft}}||\mathrm{PSF}\rangle\rangle\odot\hat{U}_{\mathrm{qft}}|\mathrm{S}\rangle).
\end{eqnarray}
Evaluating this tensor network operation for a finite number of qubits $n=2\log_{2}(L)$ is exactly equivalent to compute $I_\mathrm{img}$ finite grid of size $L\times L$ by matrix multiplication and FFT.  An implementation is shown in the plot below (see \efig{fig_convolution}) for the case of the circular aperture. \textcolor{black}{We use as input the picture used in the beginning of this article, and put it through the optical system \eref{convolution_eq} for several sizes.}
\textcolor{black}{We plot the run times versus a classical FFT implementation for comparisons for increasing approximation error $\epsilon$. Here we choose $\epsilon$ as the same cutoff in precision for both input MPS: the input image and the input field, while the QFT MPO and the multiplication run in machine precision. Also again, run times take into consideration the encodings of both MPS.} The difference in scaling is similar to the simpler case of the PSF and the same conclusions can be drawn. 
\textcolor{black}{The dominant contribution in the entanglement of the entire chain comes from the input image approximation, but it might be useful to point out that, a time advantage can still remain even if the image MPS is no longer in the compressive regime $\xi>1$, since the rest of the contractions are very low rank. The entire computation can still be computationally efficient at large-scale even if the image MPS alone might be widely inefficient.}

In this formalism, the entire image formation chain can be computed efficiently, with varying levels of complexity depending on the source image, aperture, and aberrations. In addition, other components of realistic systems, such as polarization, detector modulation functions and even noise can also be incorporated. In particular, incoherent imaging and statistical optics are easily generalizable with these methods. As the name suggests, Fourier optics is heavily based on the fact that classical optics becomes computationally faster in Fourier space as convolutions become multiplications. In the following section we will exploit this feature in more depth and apply it to other classical optical propagations. 

\section{Wave optics and quantum mechanics}
Interestingly, many other optical propagators might be good candidates for finite-rank MPO approximations, as they often respect strict conditions of smoothness, locality and symmetry. Previously we have considered the far-field regime described by the Fraunhofer propagation (the Fourier transform), a similar analysis can be generalized to the near-field regime, namely the Fresnel propagation which is valid for small distance of propagation in the paraxial approximation, and more generally to the free propagation at any distance.

\subsection{Free optical propagation}
 In full generality, we are considering here the angular spectrum transfer function which can accurately simulate propagation at any distance $z$. It describes the free homogenous propagation of a monochromatic optical field $E_z(x, y)$ over a distance $z$ in the frequency domain. It is given by
\begin{equation}\label{eq_astf}
\tilde{T}_z(f_x, f_y) = \exp\left( 2i\pi z \sqrt{ \frac{1}{\lambda^2} - f_x^2 - f_y^2 } \right),
\end{equation}
where $f_x, f_y$ are the spatial frequencies in the $x$ and $y$ directions, $\lambda$ is the wavelength of the wave and $z$ is the propagation distance. This propagator is the solution of the Green function of the following Helmholtz equation
 \begin{equation}\label{Helmholtz}
\left(\frac{\partial^2}{\partial z^2}+
\nabla_\perp^2\right) E_z +
k^2 E_z=0.
\end{equation}
 In the following, we fix  $\lambda= 633nm$ and $\Delta x=\Delta y=10 \mu m$ for all numerical simulations. 
    \begin{figure}[ht!]
\includegraphics[scale=0.08]{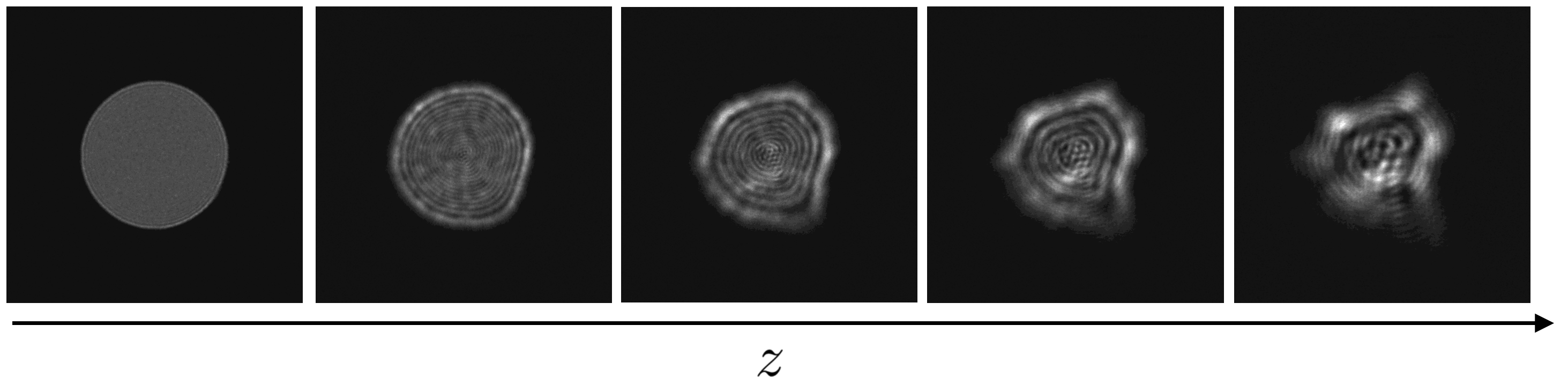} 
	\caption{Angular spectrum propagation \eref{eq_astf} of an aberated wave-front for increasing distance $z$.}
	\label{fig_propa}
\end{figure}

\begin{figure}[ht!]
\includegraphics[scale=0.17]{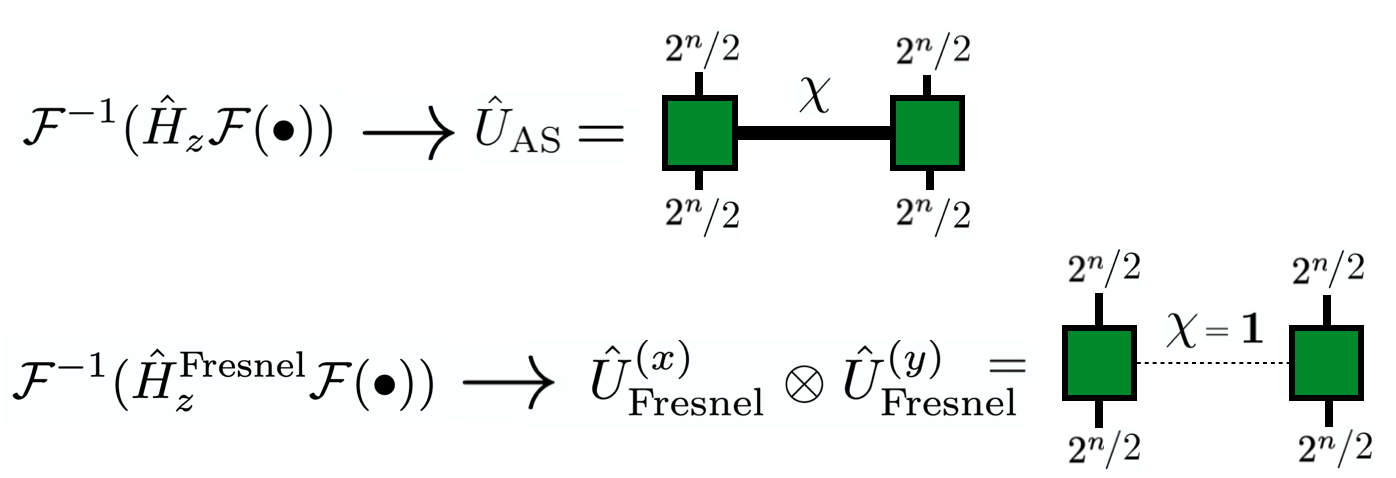} 
	\caption{Spatial separability of the optical propagators in tensor notation.}
	\label{fig_AS_fresnel}
\end{figure}

\begin{figure}[ht!]
\includegraphics[scale=0.14]{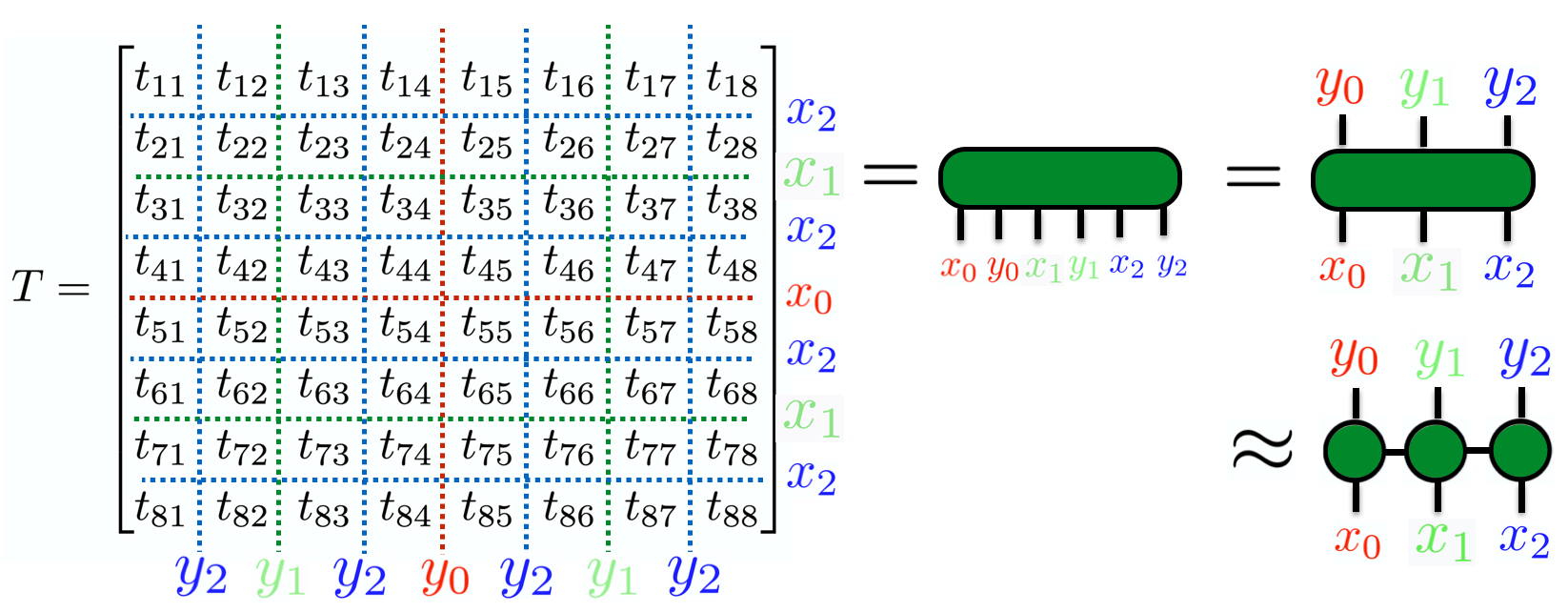} 
	\caption{Interleaved construction of a matrix product operator from a matrix. The construction is similar to the quantics MPS, only the placement of the indices are different.}
	\label{MPO_construction}
\end{figure}

\begin{figure}[ht!]
\includegraphics[scale=0.27]{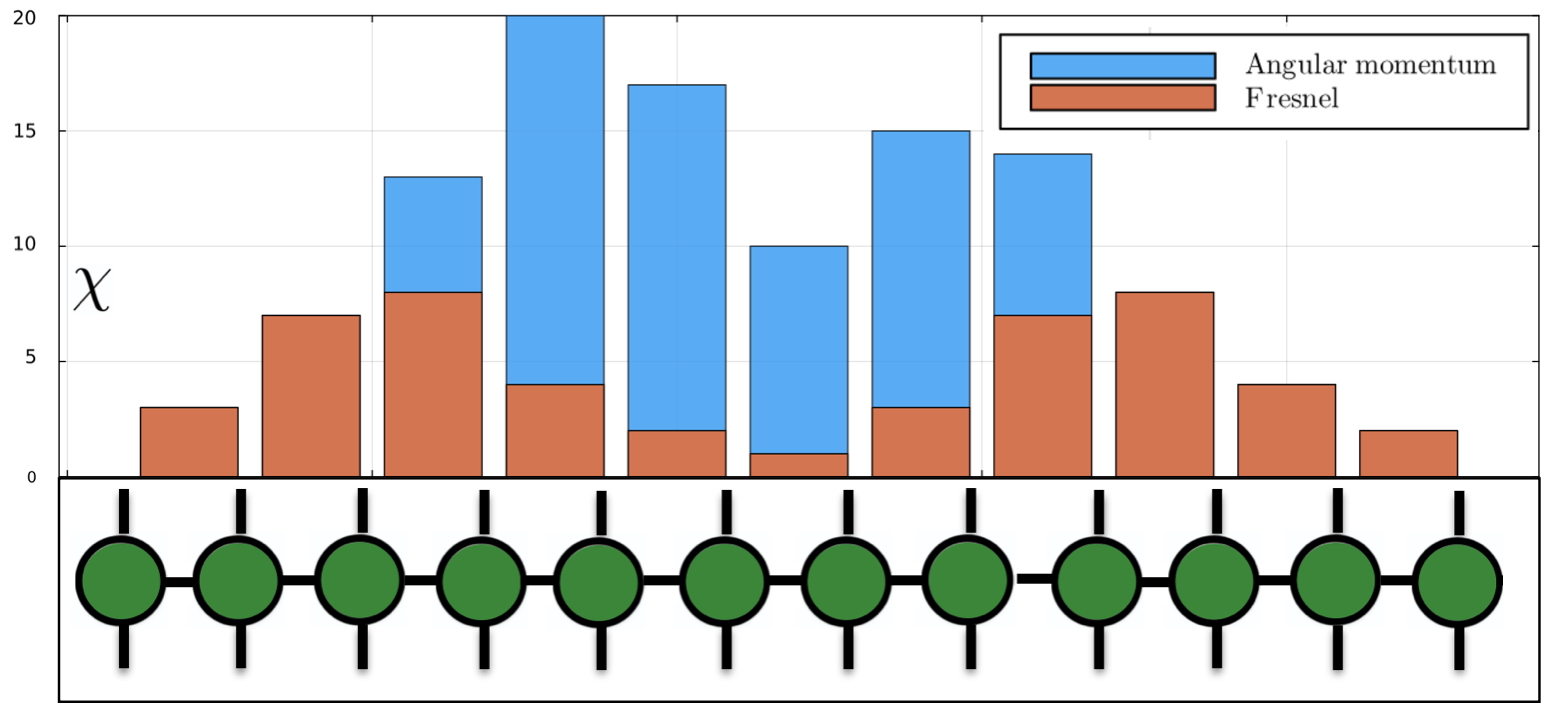} 
	\caption{Bond dimensions $\xi$ along a chain of 12 qubits for the $2d$ Fresnel and $2d$ ASTF matrix product operators in a fused representation \textcolor{black}{for $z=0.1 m$}. In this setting, the left half of the chain corresponds to the $x$-component and the right half corresponds to the $y$-component.}
	\label{fig_mpo_fresnel}
\end{figure}

To simulate wave propagation, the field is transformed into the frequency domain 
$\tilde{E}_0(f_x, f_y) = \mathcal{F}\{ E_0(x, y) \}$.
The field at distance $ z $ is then given by $\tilde{E}_z(f_x, f_y) = \tilde{E}_0(f_x, f_y)\cdot T_z(f_x, f_y)$. For evanescent components, where $f_x^2 + f_y^2 > 1/\lambda^2$, the square root becomes imaginary. The real space propagator that transform $E_0$ into $E_z$ is then described a unitary operator (see \efig{fig_AS_fresnel}).
  \textcolor{black}{As a warm-up exercise, before tackling the more difficult task of systematically constructing low-rank matrix product operators for optical propagators, we first employ a brute-force method (computationally infeasible for large 
$n$) to diagnose the structure of these operators. In \efig{MPO_construction}, we show how to construct MPO by SVD in a similar fashion we used for MPS previously.}
 The propagator \efig{fig_AS_fresnel} described as a MPO $\hat{U}_{AS}$, can not be written as separable operator $U^{(x)}\otimes U^{(y)}$ due to the the square root in \eref{eq_astf} causing couplings between the frequencies.  The amount of coupling (entanglement) can nevertheless be quantified by organizing the transfer function into a matrix product operator with a fused representation and looking at the rank between the two halves of the operator (see \efig{fig_mpo_fresnel}).
We observe that the full angular spectrum operator admits a high degree of half-chain entanglement that grows with the propagation distance $z$.
Assume that $ f_x^2 + f_y^2 \ll \frac{1}{\lambda^2} $. Let $f^2 = f_x^2 + f_y^2$, $\epsilon = \left( \frac{2\pi}{k} \right)^2 f^2$ and $\sqrt{k^2 - (2\pi)^2 f^2} = k \sqrt{1 - \epsilon}$. Using a Taylor expansion $\sqrt{1 - \epsilon} \approx 1 - \frac{\epsilon}{2}+...$, the propagator can be written as a product
\begin{equation}
\tilde{T}_{z}^{\mathrm{Fresnel}}(f_x, f_y)=\mathrm{e}^{izk}\tilde{T}_z(f_x)\tilde{T}_z(f_y),
\end{equation}
with $\tilde{T}_z(f) = \exp(-i \pi \lambda z f^2)$ and $T_z(x) = \frac{\mathrm{e}^{ikz}}{i \lambda z} \exp\left( \frac{ik}{2z}x^2 \right)$. We have shown that under the paraxial approximation $ f_x^2 + f_y^2 \ll \frac{1}{\lambda^2} $, the angular spectrum transfer function becomes separable (see \efig{fig_AS_fresnel}). Using the operator notations we have $\hat{U}_{\mathrm{Fresnel}}=\hat{U}^{(x)}_{\mathrm{Fresnel}}\otimes\hat{U}^{(y)}_{\mathrm{Fresnel}}$ and $\lim_{z\rightarrow 0}\hat{U}_{AS}=\hat{U}^{(x)}_{\mathrm{Fresnel}}\otimes\hat{U}^{(y)}_{\mathrm{Fresnel}}$.

 Numerical simulations (see \efig{fig_mpo_fresnel}) suggest that low-rank representations of these operators might exist within the range of parameters we are working in.  
\begin{figure}[!ht!]
\includegraphics[scale=0.3]{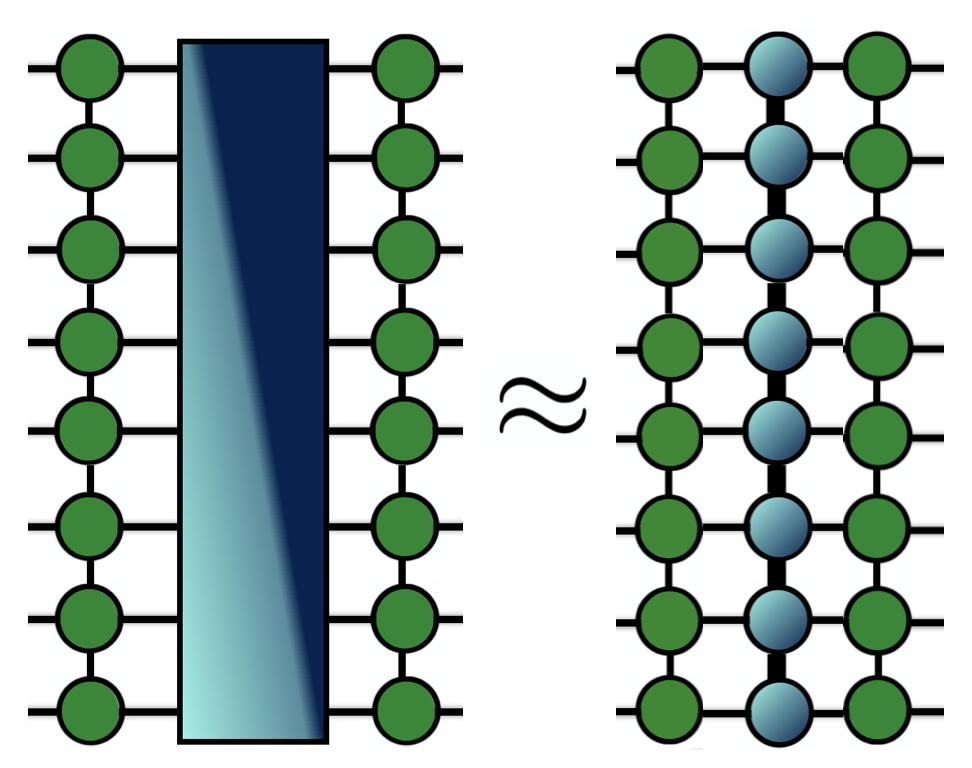} 
	\caption{MPO representation of the optical propagation as a product $\hat{U}_{\mathrm{QFT}}^\dagger \hat{U}(z) \hat{U}_{\mathrm{QFT}}$, with $\hat{U}(z)=\mathrm{exp}(-iz \hat{H})$ being a diagonal operator.}
	\label{fig_mpo_mpo}
\end{figure}
Both transfer functions are unitary and diagonal in Fourier space, which alone motivates us to construct discrete diagonal MPO representations by interpreting them as the unitary evolution (see \efig{fig_mpo_mpo}) of free fermions on $1d$ lattices.  In the following section, we focus on the $1d$ Fresnel propagator and systematically construct a low-rank MPO representation, drawing on an analogy with the evolution of a quantum free particle and its discretized spin-chain equivalent.
 
\subsection{Fresnel propagator and the free quantum particle}

As alluded to earlier, readers familiar with quantum mechanics 101 will readily notice the similarities between the Fresnel propagator and the unitary propagator of a non-relativistic quantum particle of mass 
$m$. For a free particle at position $\vec{r}$ and momentum $\vec{p}$, the Hamiltonian is  
\begin{equation}
    \hat{\mathcal{H}}_{\mathrm{free}} = \frac{\hat{p}^2}{2m}
    = \frac{\hbar^2 \hat{k}^2}{2m},
\end{equation}  
where $\hat{p} = \hbar \hat{k}$ and $\hat{k} = -i\,\partial_r$ in the position representation.  
The time-evolution operator (propagator) is  
\begin{equation}
    \hat{U}_{\mathrm{free}}(t)
    = \exp\!\left(-\frac{i}{\hbar}\,\hat{\mathcal{H}}_{\mathrm{free}}\,t\right).
\end{equation}  
Momentum eigenstates $\ket{k}$ satisfy $\hat{k}\,\ket{k} = k\,\ket{k}$. Acting on such a state, the propagator yields in $2d$
\begin{equation}
    \hat{U}_{\mathrm{free}}(t)\,\ket{k}
    = \mathrm{e}^{-i E_k t/\hbar}\,\ket{k},
    \qquad
    E_k = \frac{\hbar^2 (k_x^2+k^2_y)}{2m}.
\end{equation}  
Equivalently, the momentum-space matrix elements are  
\begin{equation}
    \bra{k} \hat{U}_{\mathrm{free}}(t) \ket{k'}
    = \mathrm{e}^{-i E_k t/\hbar}\,\delta(k-k').
\end{equation}  
Thus, although $\hat{k}$ is an operator in position space, in the momentum basis it is represented by its eigenvalue $k$, and the Hamiltonian (and therefore the propagator) becomes diagonal.  Similarly, the $2d$ Fresnel transfer function describing paraxial wave propagation for wavelength $\lambda$ over a transverse distance $z$ is
\begin{equation}
    \hat{U}_{\mathrm{Fresnel}}(k,z) = \exp\left(-i \frac{\lambda (k_x^2+k_y^2)}{4\pi} z\right).
\end{equation}
This can be interpreted as a unitary evolution generated by the Hamiltonian
\begin{equation}
    \hat{\mathcal{H}}_{\mathrm{Fresnel}} = -\frac{\lambda}{4\pi} \big(\partial_x^2+\partial_y^2\big),
\end{equation}
so that $\hat{U}_{\mathrm{Fresnel}}(z) = e^{-i z \hat{\mathcal{H}}_{\mathrm{Fresnel}}}$. Thus, Fresnel propagation is formally analogous to the quantum evolution of a free-particle in one dimension, with the parameter identification
\begin{equation}
    z \leftrightarrow t, \quad \frac{\lambda}{2\pi} \leftrightarrow \frac{\hbar}{m}.
\end{equation}
This similarity comes from the very well known equivalence between the Schrödinger equation and paraxial Helmholtz equations
\begin{equation}
\left(i\partial_z+\frac{\lambda}{4\pi}\partial_r^2\right)E= 0  \leftrightarrow \left(i\partial_t+\frac{\hbar}{2m}\partial_r^2\right)\Psi= 0. 
\end{equation}
The addition of a potential $V(x)$ in the former would imply a heterogenous refraction index $n(x)$ in the later. \textcolor{black}{We shall use the time representation (rather than $z$) in the following sections to better match standard quantum hamiltonian evolution formalism.}  As said earlier, the Fresnel/free particle propagator has, like the Fourier transform, a separable tensor product structure in space, so we can focus on its $1d$ discretization in the following.

\subsubsection{Discretization of the Hamiltonian}

Discretizing the $1d$ space into a lattice with $L$ sites and spacing $a$, the second derivative operator is approximated by the discrete Laplacian
\begin{equation}
    \frac{\partial^2 \Psi}{\partial x^2}\Big|_j \approx \frac{\Psi_{j+1} - 2 \Psi_j + \Psi_{j-1}}{a^2}.
\end{equation}
This corresponds to a tight-binding Hamiltonian on a one-dimensional chain  
\begin{equation}\label{tight_binding}
    \hat{H} = -J \sum_{j=1}^{L-1} \left( \ket{j}\bra{j+1} + \ket{j+1}\bra{j} \right) + \mu \sum_{j=1}^L \ket{j}\bra{j},
\end{equation}
with hopping strength $J= \lambda/4\pi a^2$ and onsite potential $\mu = \frac{\lambda}{2}$.
Using fermionic operators $c_j, c_j^\dagger$ for each site, with periodic boundary conditions $c_{L+1} \equiv c_1$, the Hamiltonian reads
\begin{equation}
    \hat{H} = -J \sum_{j=1}^{L-1} \left( c_j^\dagger c_{j+1} + c_{j+1}^\dagger c_j \right) + \mu \sum_{j=1}^L c_j^\dagger c_j.
\end{equation}
Fourier transforming to momentum space diagonalizes the Hamiltonian  
\begin{equation}
    \hat{H} = \sum_k \epsilon(k) c_k^\dagger c_k,
\end{equation}
with dispersion relation $\epsilon(k) = -2J \cos k + \mu, \quad k \in [-\pi, \pi]$. Expanding near $k=0$ gives  $\cos k \approx 1 - \frac{k^2}{2} \rightarrow \epsilon(k) \approx -2J + J k^2 + \mu$.
Ignoring the constant shift $(-2J + \mu)$, the dispersion is approximately quadratic $\epsilon(k) \approx J k^2$, which matches the continuous Fresnel dispersion by identifying $Ja^2 \leftrightarrow \frac{\lambda}{4\pi}$. In the following, we choose to keep the analysis at the level of the lattice with the oscillating dispersion as we could always take the parabolic limit in the end.

\subsubsection{Hamiltonian in the qubit basis}

Having written the Fresnel transform as the unitary propagation of a quantum Hamiltonian of $L$ hopping particles on a lattice, we can proceed to compress the Hilbert space acting on $n=\log_2{L}$ qubits. Consider an $n$-qubit computational basis labeled by bitstrings $\mathbf{s} = (s_1, s_2, \ldots, s_n)$ where each $s_j \in \{0,1\}$. Define the number $x_{\mathbf{s}} = \sum_{j=1}^n s_j 2^{-j} \in [0,1)$, we then need to express the dispersion relation of the diagonalized Hamiltonian such that $\hat{H}$ remains diagonal in the computational basis. Any diagonal hamiltonian (or operator for that matter) can be expanded in Pauli-strings of any size
\begin{equation}
\hat{H}= \sum_{S \subseteq \{1,\ldots,n\}} c_S \prod_{j \in S} \hat{Z}_j,
\end{equation}

where the coefficients are real and expressed as
\begin{eqnarray}\label{eq  coeff_trace}
c_S &=& \frac{1}{2^n} \mathrm{Tr}\left(\hat{H} \prod_{j \in S} \hat{Z}_j\right)\nonumber\\
&=& \frac{1}{2^n} \sum_{\mathbf{s} \in \{0,1\}^n} \epsilon(\mathbf{s}) (-1)^{\sum_{j \in S} s_j}.
\end{eqnarray}
This decomposition is the Walsh–Hadamard transform (which is a generalized  Fourier transform over the abelian group $\mathbb{Z}_2^n$),
it can be computed very efficiently for large-system with the fast Walsh–Hadamard transform (see for example \cite{georges2025pauli} and references within). Generally speaking, diagonal operators are crucially important in quantum computing as they are key elements of many quantum algorithms \cite{bullock2003smaller,bullock2008asymptotically,welch2014efficient,nakata2014diagonal}.


In our specific case $\epsilon(\mathbf{s}) = \cos(2\pi x_{\mathbf{s}})$, a standard calculation ({\it cf.} appendix B) gives the coefficients $c_S \in \mathbb{R}$ equal to
\begin{equation}\label{eq_cs}
c_S = \frac{1}{2^n} \operatorname{Re} \left[ \prod_{j=1}^n \left(1 + (-1)^{\mathbf{1}_{j \in S}} \mathrm{e}^{i \theta_j} \right) \right],
\end{equation}
where we defined $\theta_j   = \frac{2\pi}{2^j}$ and $S_{1}=\hat{Z}_i$, $S_{2}=\hat{Z}_i\hat{Z}_j$, $S_{3}=\hat{Z}_i\hat{Z}_j\hat{Z}_k$ for $i,j,k \in [1,n]$ and so on. There is an exponentially large number of terms in the Pauli decomposition, but hopefully a finite number of them are contributing significantly to the sum. By separating the product over indices in $S$ and not in $S$, we can write $c_S$ as
\begin{equation}
\boxed{
c_S = \frac{1}{2^n} \operatorname{Re} \Big( \gamma_n \prod_{j \in S} x_j \Big),}
\end{equation}
with $\gamma_n=\frac{e^{i \pi \left(\frac{1}{2} - \frac{1}{2^n}\right)}}{\sin\left(\frac{\pi}{2^n}\right)}
$ and $x_j = -i \tan(\theta_j/2)$.
 As $n$ goes to infinity the Hamiltonian is made of absolutely convergent coefficients $c_S$.
This hamiltonian, although made of Pauli-strings of any length at any distance, is effectively highly-local as $n$ grows.
Knowing that the two-body interactions are essentially limited to a few neighbors, what are the contributions coming from the longer strings of operators and what are their dependences to their relative size. Following this path, we traded the nearest-neighbors fermionic hamiltonian \eref{tight_binding} with $L$ sites with a $n$-qubit system with exponentially decreasing interactions. This result relies heavily on the analyticity of the dispersion relation and is not expected to give a similar result in general\footnotemark\footnotetext{The Walsh-Fourier version of the Paley–Wiener theorem states that, for any finite abelian group, if $\epsilon(\mathbf{s})$ is analytic in a strip around the real line, then its Walsh–Fourier coefficients decay exponentially fast while smoothness alone would result in a polynomial decrease at best \cite{rudin2017fourier}.}.  
 \begin{figure}[!ht!]
\includegraphics[scale=0.24]{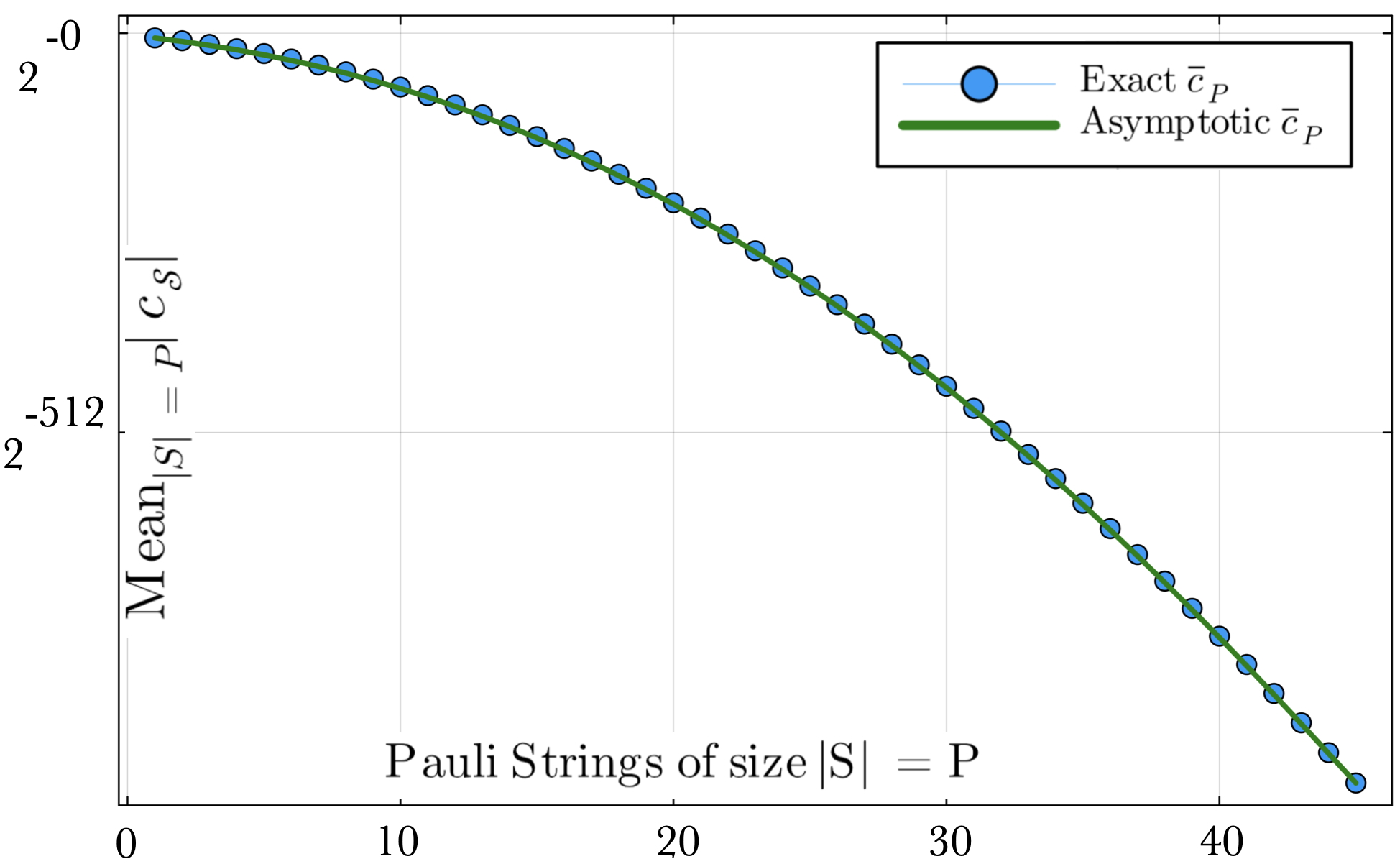} 
	\caption{Exponential decay of the mean amplitudes of Pauli strings of size $P$. The exact formula is given by \eref{exact_formula} and the asymptotic by \eref{asymptotic_c}.}
	\label{pauli_dec_strings}
\end{figure}
Let us be more precise and compute the mean amplitude of Pauli-strings of size $|S|=P$
\begin{equation}
\overline{c}_P = \frac{1}{\binom{n}{P}} \sum_{|S|=P} c_S.
\end{equation}
It is possible to compute the asymptotical behavior of the average amplitude of Pauli-strings. Indeed we can write $\overline{c}_P$  as the coefficient of the generating function of all Paul-strings of size $P$.
\begin{equation}\label{exact_formula}
\overline{c}_P = \frac{1}{2^n \binom{n}{P}} \operatorname{Re} \Big[ \gamma_n e_P(x_1, \dots, x_n) \Big].
\end{equation}
The \emph{elementary symmetric polynomial of degree $P$} in $n$ variables $x_1, \dots, x_n$, well known in the context of the fractional quantum Hall effect \cite{macdonald1998symmetric} is defined as
\begin{equation}
e_P(x_1, \dots, x_n)   = \sum_{1 \le j_1 < \dots < j_P \le n} x_{j_1} x_{j_2} \dots x_{j_P},
\end{equation}
and $e_0   = 1$ \footnotemark\footnotetext{Actually, all the moments can be written using those polynomials, giving access to $U(t)$ as an exact moment expansion.}. Thanks to the convergence properties of $c_S$, we can show (more details in the appendix B) that the asymptotic behavior with respect to $P$ is
\begin{equation}\label{asymptotic_c}
\overline{c}_P \sim 2^{-P(P+1)/2} \left(\frac{\pi}{N}\right)^P P! \quad \text{for } P \ll n.
\end{equation}
This decay is super-exponential in $P$ and it shows that the amplitudes of Pauli strings do not contribute at all past a certain string length. We plot in \efig{pauli_dec_strings} this asymptotic behavior and compare it to the exact formula computed with the fast Walsch-Hadamard transform.

This asymptotic behavior is key to select a polynomially long sequence of gates that still retains machine precision and approximates the diagonal operator with an efficient MPO. If one truncates the Hamiltonian to include only Pauli strings of length $P_m$, the truncated unitary reads $\hat{U}_\mathrm{trunc}(t) = e^{-i t \hat{H}_\mathrm{trunc}}$, with $\hat{H}_\mathrm{trunc} = \sum_{|S|\le P_m} c_S \hat{\mathrm{P}}_S$.  
Since all terms commute, the truncation error in operator norm can be bounded by the norm of the discarded tail Hamiltonian 
\begin{equation}
\|\hat{U}(t) - \hat{U}_\mathrm{trunc}(t)\| \le \| t \hat{H}_\mathrm{tail} \|_\mathrm{op} \le t \sum_{|S|>P_m} |\overline{c}_S|.
\end{equation}
Using the expression for $\overline{c}_P$, we get
\begin{equation}\nonumber
\|\hat{U}(t) - \hat{U}_\mathrm{trunc}(t)\| \lesssim t \sum_{P=P_m+1}^{n} \binom{n}{P} 2^{-P(P+1)/2} \left(\frac{\pi}{N}\right)^P P!.
\end{equation}
Because the super-exponential decay $2^{-P(P+1)/2}$ dominates the combinatorial growth $\binom{n}{P}$ for moderate $P$, the truncation error decreases extremely rapidly with $P$, justifying the efficient approximation of $\hat{U}(t)$ by a sum of only short Pauli strings and giving access to a compact MPO representation using a polynomial $\mathcal{O}(n^{P_m})$ number of terms.

\subsubsection{Unitary evolution,  quantum circuit and MPO}

\begin{figure}[!ht!]
\includegraphics[scale=0.088]{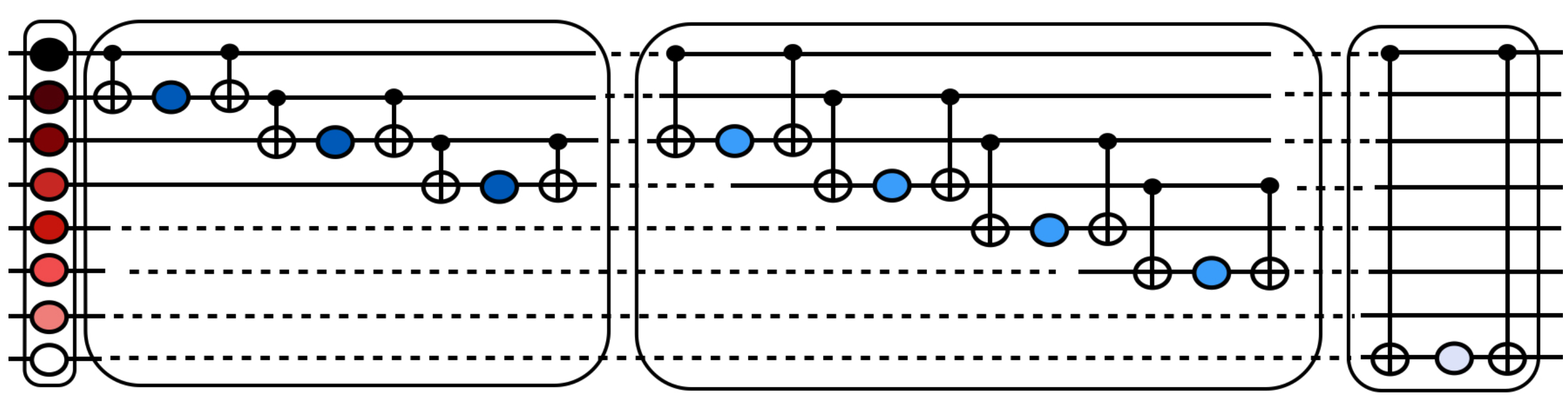} 
	\caption{Circuit representing the unitary evolution \eref{eq  U-product_1} made of only CNOT and $R_z$ gates.}
	\label{fig_circuit}
\end{figure}

Let us define the string operators $\hat{\mathrm{P}}_S=\prod_{j \in S} \hat{Z}_j$ such that $\hat{D}=\sum_{S\subseteq\{1,\dots,n\}} c_S \,\hat{\mathrm{P}}_S$. Each $\hat{\mathrm{P}}_S$ is Hermitian and unitary $\hat{\mathrm{P}}_S^\dagger = \hat{\mathrm{P}}_S$ and $\hat{\mathrm{P}}_S^2 = \hat{I}$ and different Pauli-$\hat{Z}$ strings commute, $[\hat{\mathrm{P}}_S,\hat{\mathrm{P}}_{S'}]=0\quad\text{for all }S,S'$.
Because the terms in the hamiltonian commute pairwise, the exponential of the sum factorizes exactly into the product of exponentials, the unitary operator becomes
\begin{equation}
\hat{U}(t)=\prod_{S\subseteq\{1,\dots,n\}} \mathrm{e}^{-i c_S t\, \hat{\mathrm{P}}_S}.
\end{equation}
Moreover, each factor $\mathrm{e}^{-i\omega \hat{\mathrm{P}}_S}$ has a closed form. Using the power-series definition of the exponential and separating even/odd powers,
we obtain the elementary trigonometric form for the exponential operator
\begin{equation}\label{eq  single-factor}
\mathrm{e}^{-i\omega \hat{\mathrm{P}}_S} = \cos(\omega)\hat{I}-i\sin(\omega)\hat{\mathrm{P}}_S .
\end{equation}
Applying this with $\omega=c_S t$ gives each factor explicitly, therefore the exact evolution is
\begin{equation}\label{eq  U-product_1}
\boxed{
\hat{U}(t)=\prod_{S}\big[\cos(\omega)\hat{I} - i\sin(\omega)\,\hat{\mathrm{P}}_S\big].}
\end{equation}
This formula allows to construct a quantum circuit made only of CNOT and $R_z(2\omega)$ gates with coefficients $\{\omega\}$ given by \eref{eq_cs}.
In \efig{fig_circuit}, we illustrate a quantum circuit, with each $n$-qubit gates corresponding to a $n$-Pauli string in the decomposition.
Because of the commutativity of the Pauli strings, the results are independent of the order of gates.
The resulting matrix product operator is made by sequentially absorbing all the local tensors into an initial identity MPO. Numerically \footnotemark\footnotetext{We can directly use native $\mathrm{ITensor}$ functions  $U = \operatorname{apply}\!\bigl( \operatorname{ops}(\texttt{gates}, \texttt{qsites}),  
\operatorname{MPO}(\texttt{qsites}, "I"; \,\texttt{cutoff}=10^{-6}) \bigr)$ where $\texttt{gates}$ is a list of gates} , we select gates based on their amplitude and a cutoff at $10^{-6}$ is chosen, which is garanties a precision below $10^{-10}$ for the unitary.   
\begin{figure}[!ht!]
\includegraphics[scale=0.22]{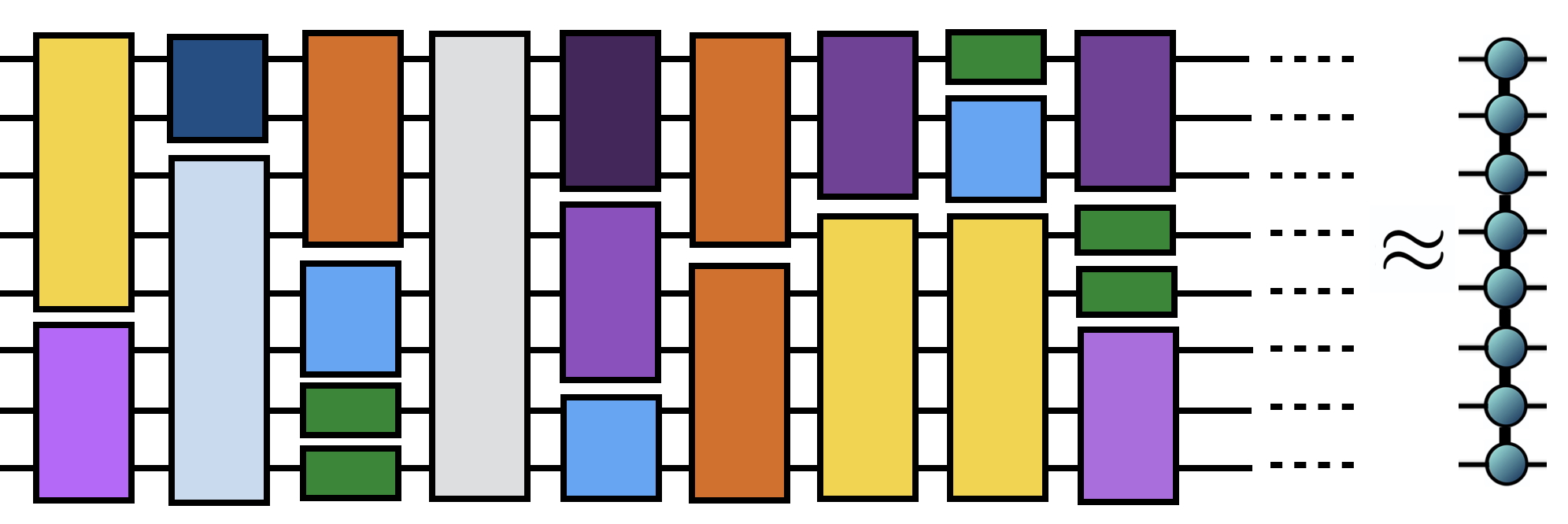} 
	\caption{Quantum circuit (\efig{fig_circuit}) as a big tensor network and associated matrix product operator. The blocks of lengths $P$ correspond to $P$-qubit gates. The order is irrelevant due to commutativity.}
	\label{fig_exact_circuit}
\end{figure}
In \efig{fig_chi_max}, we plot the maximal rank of the MPO up to $n=20$ and show that the entanglement saturates at a finite value when $n$ grows which suggests that this diagonal operator is of finite-rank despite being constructed of an exponential number of Pauli-strings.
\begin{figure}[!ht!]
\includegraphics[scale=0.203]{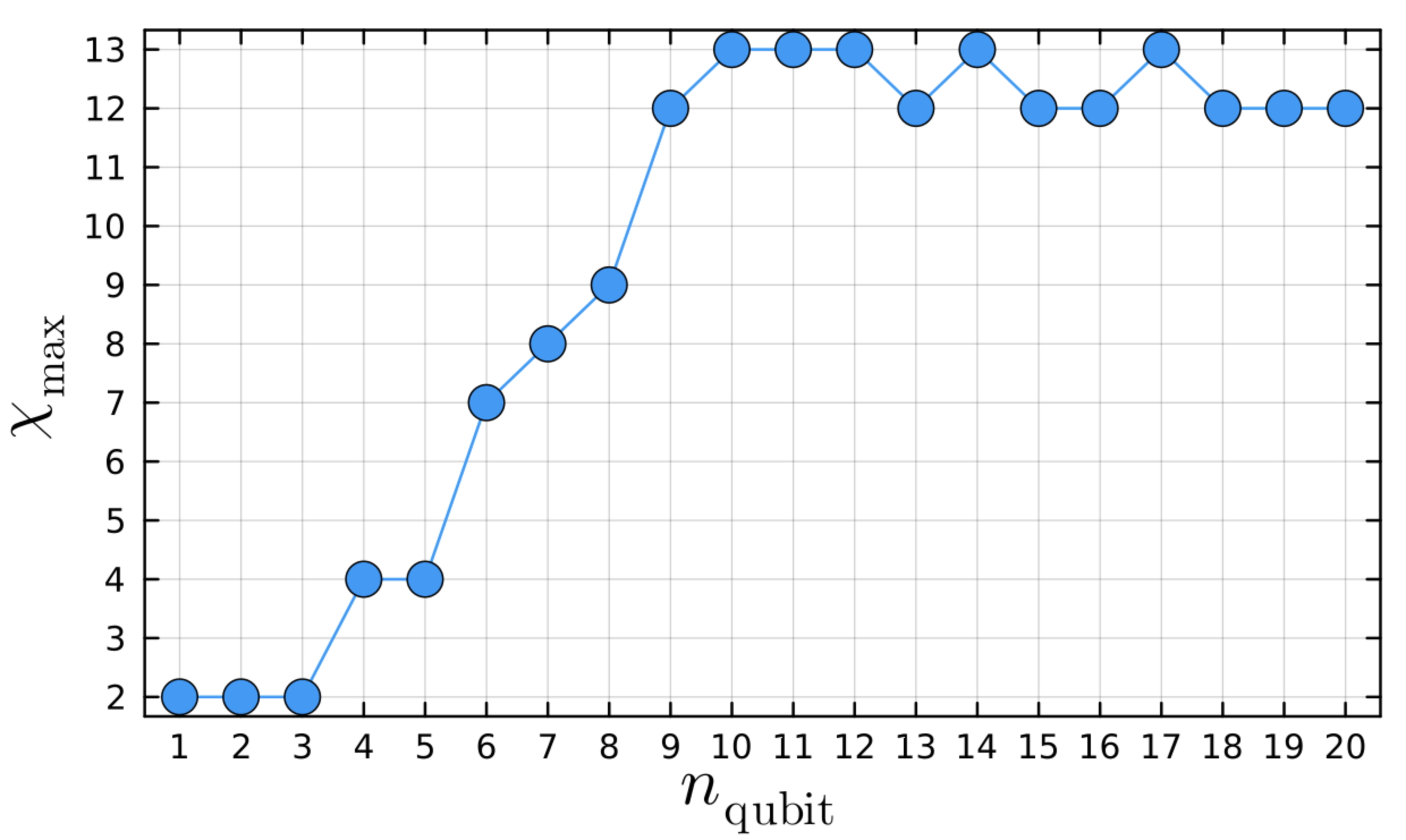} 
	\caption{Maximal rank of the MPO representing \efig{fig_exact_circuit} when adding up all gates below $10^{-6}$. The bond dimension $\chi_\mathrm{max}$ of the MPO saturates (for $t=1$) quickly when $n$ grows, indicating a low-rank structure.}
	\label{fig_chi_max}
\end{figure}
Similarly to the Fourier transform MPO, we have successfully showed that the Fresnel transform in Fourier space also admits a low rank structure in Fourier space. The real-space operator is also low-rank since it is a product of low-rank MPO (see \efig{fig_mpo_mpo}). The numerical method developed here is very generic and is expected to work with any diagonal operator. Of course the maximal rank of the MPO will grow with the complexity of the function $\mathbf{\epsilon}(\mathbf{s})$, but can always be controlled by truncating the terms in the Pauli expansion. As long as the behavior of $\overline{c}_p$ is sufficiently decreasing, the quantum circuit and the MPO associated should stay numerically tractable for large value of $n_\mathrm{qubit}$. Additionally, more subtle and efficient ways, such as recursive decomposition into multiplexed 
$R_z$ gates \cite{shende2005synthesis,bullock2003smaller} allows to construct the MPO using $\mathcal{O}(n)$ gates. In the next section, we will introduce another method that is more mathematically explicit and offers some advantages in particular circumstances.

\subsubsection{Closed-expression for the diagonal MPO using Bessel expansion}

\begin{figure}[!ht!]
\includegraphics[scale=0.245]{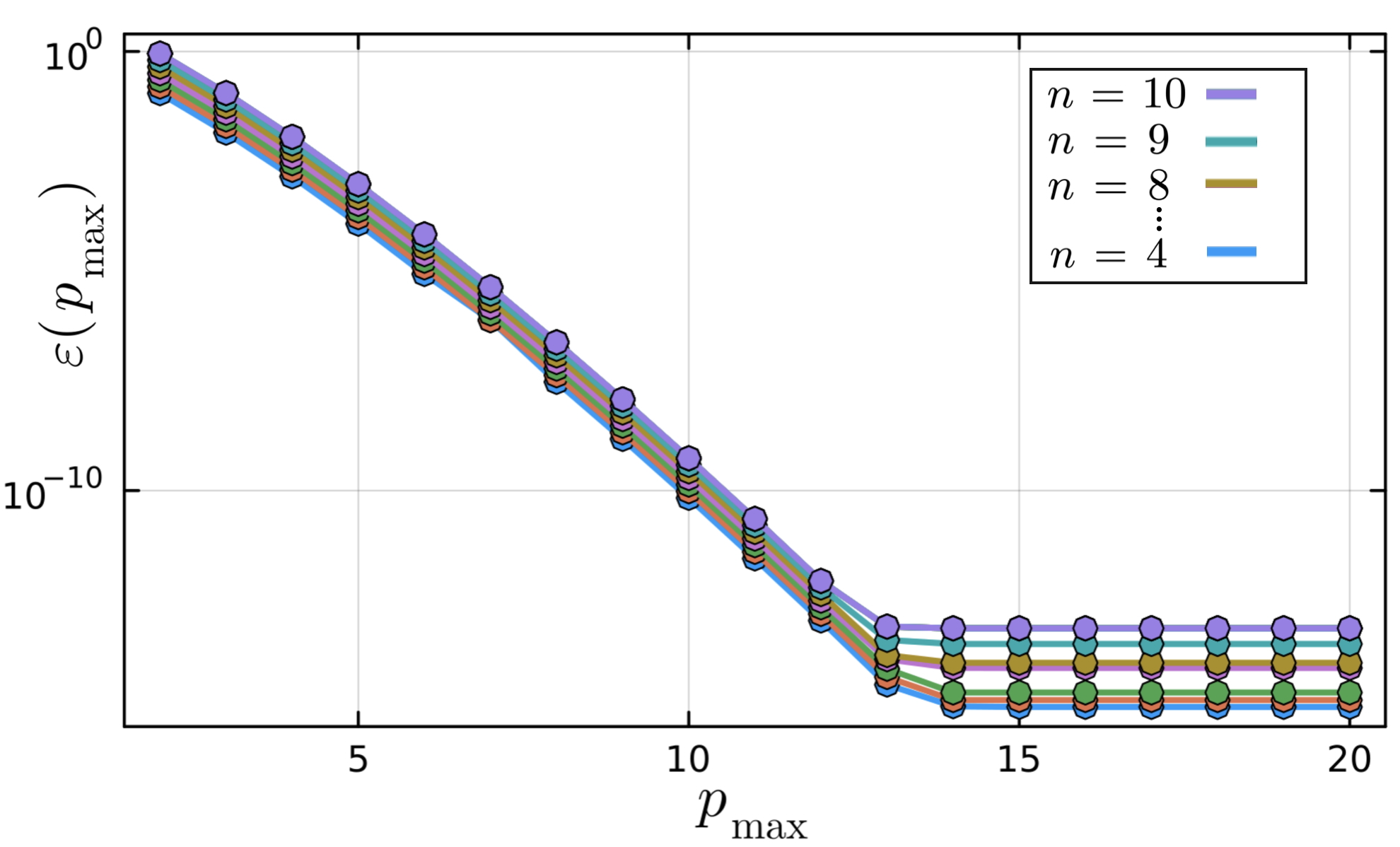} 
	\caption{Unitary matrix reconstruction error for $n=4,5,...,10$ and $t=1$. The error decreases exponentially fast and remain constant for a finite value of $k_\mathrm{max}$.}
	\label{fig_error_mpo}
\end{figure}

In this section, the goal is to write down a closed expression for a MPO representing the unitary evolution $U(t)$ by using classical tools of statistical physics of spin systems. From this point of view, the diagonal unitary evolution operator
\begin{equation}
\hat{U}(t) = \sum_{\mathbf{s} \in \{0,1\}^n} \mathrm{e}^{-i t \epsilon(\mathbf{s})} \ket{\mathbf{s}}\bra{\mathbf{s}},
\end{equation}
with $\epsilon(\mathbf{s})=\cos(2\pi\mathbf{s})$,  correspond to the partition function of a classical $1d$ system with complex Boltzmann weights. A classical trick is to use the Jacobi-Anger expansion \cite{liu2013exact,samlodia2024phase}, that states that for any $t \in \mathbb{R}$,
  \begin{align}
\mathrm{e}^{-i t \cos(2\pi\mathbf{s})} 
&= \sum_{p=-\infty}^{\infty} (-i)^p J_p(t) \mathrm{e}^{2i\pi p \mathbf{s}}\nonumber \\
&= \sum_{p=-\infty}^{\infty} (-i)^p J_p(t) \prod_{j=1}^n \mathrm{e}^{i p s_j \theta_j},
\end{align}
with $\theta_j   = \frac{2\pi}{2^j}$ and where $J_p(t)$ is the $p$-th Bessel function of the first kind. The last formula is directly a matrix product operator as a superposition of plane waves weighted by Bessel coefficients. In practice, the infinite sum over $p$ is truncated to $|p| \le p_{\rm max}$. Let $\chi = 2 p_{\rm max} + 1$ be the bond dimension, corresponding to the $p$ index.  
Define local tensors $W_j$ with physical index $s_j$ and bond indices $p_{\rm in}, p_{\rm out}$ as
\begin{equation}
[W_j]_{p_{\rm in}, p_{\rm out}}^{s_j} =
\begin{cases}
\mathrm{e}^{i p_{\rm in} \theta_j s_j}, & p_{\rm out} = p_{\rm in} \\
0, & \text{otherwise}
\end{cases},
\end{equation}
with $p_{\rm in}, p_{\rm out} \in \{-p_{\rm max},\dots,p_{\rm max}\}$. The left and right boundary vectors $|v^R\rangle,|v^L\rangle\in\mathbb{C}^{\chi}$ are defined by their components
\begin{equation}
\begin{cases}
v^{L}_{p} = (-i)^p J_p(t)\\
v^{R}_{p} = 1.
\end{cases}
\end{equation}
The bond dimension $\chi=2p_\mathrm{max}+1$ corresponds directly to the number of retained Fourier modes in the truncated expansion. Then the MPO contraction reproduces $\hat{U}(t)$ in an exact closed-form
\begin{equation}
\boxed{
\hat{U}(t) = 
\sum_{\mathbf{s}}
\big(
\langle v^L \,|\, W_1^{s_1}W_2^{s_2}...W_n^{s_n}|v^R \rangle
\big)
|\mathbf{s}\rangle\langle\mathbf{s}|.}
\end{equation}
To evaluate the convergence of the matrix product operator representation, we define the truncated expansion
\begin{equation}
\hat{U}_{p_{\max}} = \sum_{p=-p_{\max}}^{p_{\max}} (-i)^p J_p(t) 
\prod_{j=1}^n \mathrm{e}^{i k s_j \theta_j},
\end{equation}
and monitor the successive difference
\begin{equation}
\delta(p_{\max})   = \max 
\left| \hat{U}_{p_{\max}} - \hat{U}_{p_{\max}-1}\right|.
\end{equation}
\begin{figure}[!ht!]
\includegraphics[scale=0.225]{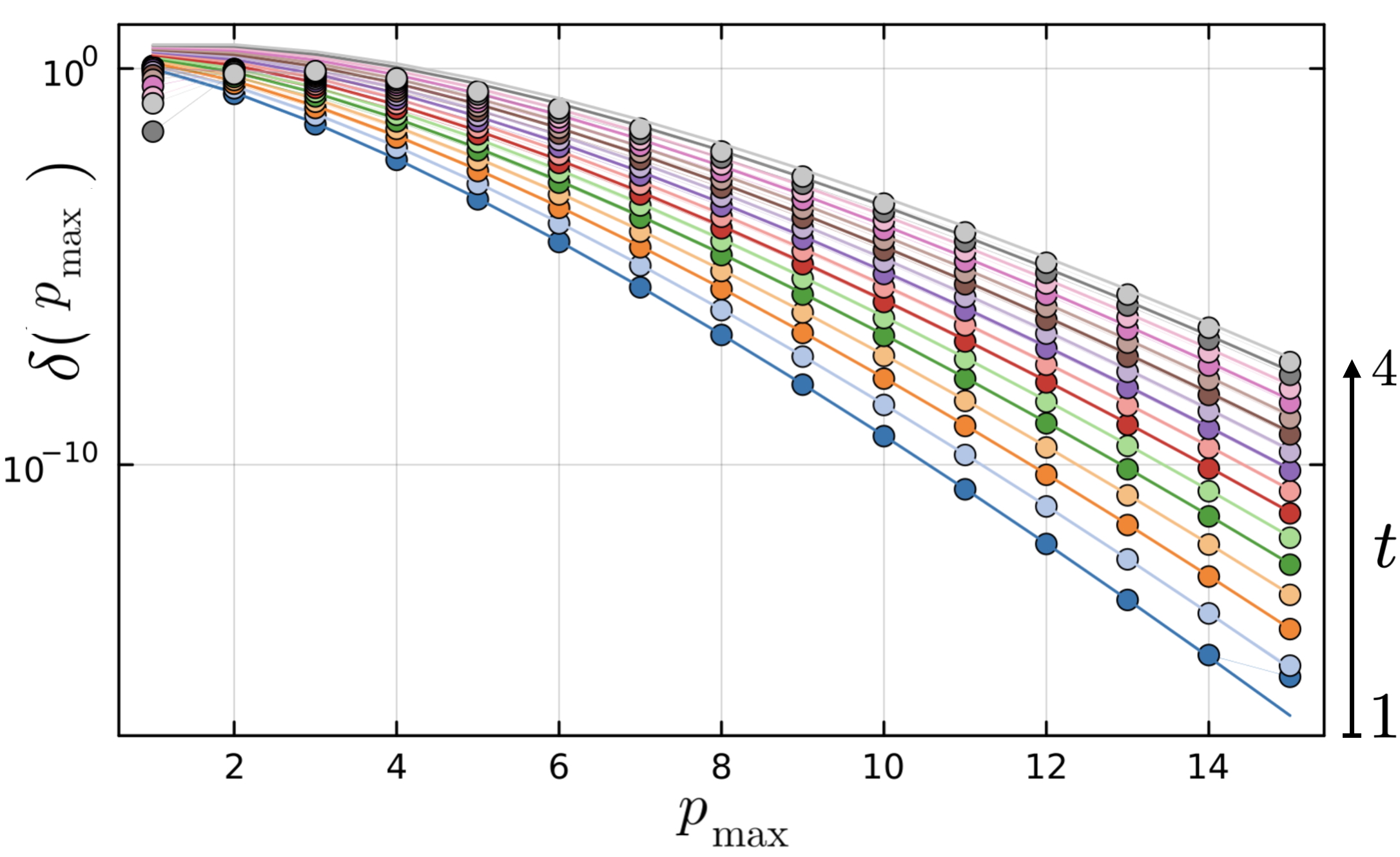} 
	\caption{Exponential convergence of the Bessel MPO for increasing values of $t\in[1,4]$, the solid lines correspond to the asymptotic estimate for $n=12$.}
	\label{fig_mpo_bessel}
\end{figure}
Since the phase factors $\prod_{j=1}^n \mathrm{e}^{i p b_j \theta_j}$ have unit modulus, the truncation error satisfies
$\delta(p_{\max}) \leq \sum_{|p| > p_{\max}} |J_p(t)|$.
Using the bound $|J_p(t)| \leq \frac{|t|^p}{p!}$, valid for all $p \in \mathbb{Z}$, we obtain the asymptotic estimate
$\delta(p_{\max})  \lesssim  
\frac{|t/2|^{p_{\max}}}{(p_{\max})!}$, which decays super-exponentially with $p_{\max}$. 
Numerical simulations confirm that for fixed $n$ the sequence $\delta(p_{\max})$ rapidly decreases as $p_{\max}$ grows, with decay rates consistent with the asymptotic Bessel bound. 
In particular, plotting $\delta(p_{\max})$ on a logarithmic scale for times $t=1,2,\ldots,10$ shows excellent agreement between the exact convergence behavior and the analytical estimate. 
In \efig{fig_error_mpo}, we show how the MPO converges toward the exact unitary matrix very quickly, at $p_{\max}=14$ the operator errors $\epsilon(p_{\max})$ are well below $10^{-10}$ for all reasonable sizes available to our computer. The same Bessel bound used above gives also an exponential scaling $\epsilon(p_{\max})\lesssim\frac{|t|^{p_{\max}}}{(p_{\max})!}$ up to floating-point roundoff.

\subsection{Angular spectrum propagator as a free relativistic quantum particle}

Now we consider the natural generalization of the free Hamiltonian in the case of relativistic quantum mechanics.
The full (non-paraxial) Helmholtz equation \eref{Helmholtz} maps directly to the $1+1d$ Klein-Gordon equation,
\begin{equation}
(\square+m^2)\Psi=0,
\end{equation}
 which is known to lead to a non-local Hamiltonian (square-root of the Laplacian) when interpreted as first-order equation. In order to obtain a proper local Hamiltonian, which can be discretize on a lattice, we can follow the Dirac interpretation of the square-root laplacian \cite{dirac1928quantum}.

This section will not be as detailed as the previous section and we will only highlight the main points of the calculations. 
We consider a $1d$ lattice with $L$ sites and spacing $a$.  
At each site $n=1,\dots,L$ we place a two-component spinor
\begin{equation}
\hat{\Psi}_i =
\begin{pmatrix}
\hat{\Psi}_{i,1} \\[4pt]
\hat{\Psi}_{i,2}
\end{pmatrix},
\end{equation}

with second-quantized fermion operators obeying
$\{\hat{\Psi}_{n,\alpha}, \hat{\Psi}_{m,\beta}^\dagger\} = \delta_{nm}\delta_{\alpha\beta}$ and
$\{\gamma^\mu, \gamma^\nu\} = 2 \eta^{\mu\nu} I, \quad \mu,\nu = 0,1,
\quad \text{with} \quad \eta = \mathrm{diag}(1, -1)$ with
$(\gamma^0)^2 = I, \quad (\gamma^1)^2 = -I, \quad \{\gamma^0, \gamma^1\} = 0$.
We discretize the derivative in the free Dirac Hamiltonian
\begin{equation}
\hat{\mathcal{H}}_\mathrm{free}^\mathrm{rel} = -i \hbar c\hat{X}\hat{\partial}_x + mc^2\hat{Z},
\end{equation}

using the symmetric finite difference
\begin{equation}
\partial_x \Psi_n   \to   \frac{\Psi_{n+1} - \Psi_{n-1}}{2a}.
\end{equation}

This yields the discrete Dirac Hamiltonian\footnotemark\footnotetext{The infamous fermion doubling problem will not pose a significant problem in our context, since we can choose one branch arbitrarily.}  
\begin{equation}
\hat{H}= -\frac{i \hbar c}{2a} \sum_n (\hat{\Psi}_n^\dagger \hat{X} \hat{\Psi}_{n+1}
+ \text{h.c.})+ m c^2 \sum_n \hat{\Psi}_n^\dagger \hat{Z} \hat{\Psi}_n,
\end{equation}
with periodic boundary conditions $\Psi_{L+1} \equiv \Psi_1$.
The equation becomes block-diagonal in momentum space
\begin{equation}
  \hat{H} = \sum_k \Psi_k^{\dagger} \hat{H}(k) \Psi_k, \qquad
  \hat{H}(k)= \frac{\hbar c}{a} (\sin k)\hat{X} + mc^2\hat{Z}.
\end{equation}
Let $\ket{u_k^{\pm}}$ denote the corresponding normalized eigenvectors.
The exact propagator in momentum space is
\begin{equation}
 U(t) = \mathrm{e}^{-i \hat{H}(k) t/\hbar}= V_k\, \mathrm{diag}\big(\mathrm{e}^{-i\epsilon_k^{+} t/\hbar},\, \mathrm{e}^{-i\epsilon_k^{-} t/\hbar}\big) V_k^{-1},
\end{equation}
where $V_k=(\ket{u_q^{+}},\ket{u_q^{-}})$ collects eigenvectors. If we select one eigensubspace, say the negative-energy branch  (and drop the - subscript), then the Hamiltonian acts as a scalar on that subspace
\begin{equation}
  \hat{H}(k)\ket{u_k} = \epsilon_k\ket{u_k}, \qquad
  \hat{U}(t)\ket{u_k} = e^{-i\epsilon_k t/\hbar}\ket{u_k},
\end{equation}
with
\begin{equation}\label{rel_dispersion}
\epsilon(k) = \sqrt{\left( \frac{\hbar c}{a} \sin(k a) \right)^2 - m^2 c^4 }.
\end{equation}
For small $k a \ll 1$,  the dispersion becomes (posing $\hbar c=1$)
\begin{equation}
\epsilon(k)=\sqrt{k^2 - m^2},
\end{equation}
recovering the continuum relation of the energy of a relativistic particle of mass $m$. Similarly to the previous section, we can identify this propagator as the angular spectrum transfer function by choosing the right parameter identification $\hbar k\leftrightarrow mc$ and $\hbar k_x\leftrightarrow p$. In the following, we choose again to stay in the lattice \eref{rel_dispersion} as we could always take the continuum limit in then end.

A similar calculation to the previous non-relativistic case can be carried out by following the same steps.
The qubit Hamiltonian writes in Fourier space
\begin{equation}
\hat{H} =\sum_{\mathbf{s}} \epsilon(\mathbf{s}) \ket{\mathbf{s}}\bra{\mathbf{s}}= \sum_{S \subseteq \{1,\ldots,n\}} c_S \prod_{j \in S} \hat{Z}_j,
\end{equation}
with 
\begin{align}
\epsilon(\mathbf{s})=\sqrt{1-\beta^2\sin^2(2\pi x_{\mathbf{s}})}.
\end{align}
Introducing the angle $\varphi_{\mathbf{s}}   = 2\pi x_{\mathbf{s}}$. Since the function $\epsilon(\mathbf{s})$ is even and $\pi$-periodic, it admits a Fourier-cosine expansion
\begin{equation}
\sqrt{1-\beta^2 \sin^2\varphi} = a_0(\beta) + \sum_{\mu=1}^{\infty} a_\mu(\beta) \cos(2 \mu \varphi),
\end{equation}
with Fourier coefficients
\begin{eqnarray}
a_0(\beta) &=& \frac{1}{\pi} \int_0^{\pi}\mathrm{d}\varphi \sqrt{1-\beta^2 \sin^2\varphi},\\
a_\mu(\beta) &=& \frac{2}{\pi} \int_0^{\pi}\mathrm{d}\varphi \sqrt{1-\beta^2 \sin^2\varphi}\, \cos(2 m \varphi) \quad \mu\ge 1\nonumber.
\end{eqnarray}
In that case the formula for the coefficient of the Pauli-string decomposition is
\begin{align}
c_S=\frac{1}{2^n}\sum_{\mu=0}^\infty a_\mu(\beta)
\mathrm{Re}\left[\prod_{j=1}^n\left(1+(-1)^{\mathbf{1}_{j\in S}}e^{i\theta_j^{(\mu)}}\right)\right],
\end{align}
where $\theta_j^{(\mu)}=\frac{\mu\,4\pi}{2^j}$. The first integral is equal to $a_0(\beta)=(1/\pi)E(\pi,\beta)$ with $E$ being the incomplete elliptic integral of the second kind and the second integral can be expressed in terms of hypergeometric functions
\begin{eqnarray}\nonumber
a_\mu= (1 + \gamma^{-1}) \frac{(-\frac12)_\mu (-\lambda^2)^\mu}{\mu!} 
\, {}_2F_1\Bigl(-\frac12, \mu - \frac12; \mu + 1; \lambda^4\Bigr),
\end{eqnarray}
for $\quad \mu \ge 1$ with $\gamma^{-1}=\sqrt{1 - \beta^2}$, and $(a)_\mu = a (a+1) (a+2) \cdots (a+\mu-1)$ and where 
$\lambda = \beta/(1 + \gamma^{-1})$.
For fixed $\mu$, the integrals converge absolutely. As $\mu\to\infty$, $a_\mu$ decays exponentially because the function $\sqrt{1 - \beta^2 \sin^2 \phi}$ is analytic in a strip around the real axis since $0 < \beta < 1$. 
Because of the exponentially decreasing behavior of the Pauli string coefficients, the MPO also admits a very low-rank representation. \textcolor{black}{We do not show the quantum unitary circuit nor the MPO implementation, as they closely mirror the Fresnel propagator.} 

In this study, the dispersion relations are analytic functions of $k$ implying exponentially fast decrease of their Hadamard coefficients. Mathematically it sounds reasonable to assert that any Hamiltonian with an analytic/holomorphic dispersion relation should admit a low-rank MPO representation. It would be very important to precisely quantify this behavior more rigorously for a larger class of functions. Alternatively, a similar Bessel-type expansion can also be found in this case, leading to a closed expression for this MPO, although we do not show details here as they do not add value to the discussion.

Additionally, it would be informative to compare the rank of the resulting MPO with that obtained from a more TCI-based implementation. Also, another quantum-inspired method, which does not rely on the Fourier transform, has been recently developed \cite{gidi2025pseudospectral} to solve quantum free evolution. Such comparisons, along with more realistic applications, will be the subject of future research. \textcolor{black}{Moreover, many other transforms that are diagonalizable in Fourier could be study within this framework. One of the most famous transform that is especially present in image processing, optics and quantum physics is the periodic Hilbert transform. In Fourier, the Hilbert operator is a simple chiral phase operator $\pm i$ which can be written $\hat{U}_{\mathrm{QFT}}^\dagger \hat{D} \hat{U}_{\mathrm{QFT}}$, with $\hat{D}=-i \hat{Z}_n$ being a rank-1 MPO. A more general analysis of Fourier operators could open new possibilities at the interface of quantum computing, filter theory and tensor networks.}

 Surprisingly enough, although we have dealt with two operators that are diagonalizable by Fourier transform, we did not construct the Fourier transform MPO by this method as its associated Hamiltonian $\hat{\mathcal{H}}=\frac{1}{2}(\hat{p}^2+\hat{x}^2)$ is not diagonalizable by Fourier transform (as $\hat{\mathcal{H}}$ is not translational invariant), but rather by the basis of discrete Hermite functions. It would be very interesting to calculate an MPO representation using this Hamiltonian formulation and compare with \cite{chen2023quantum}. 

\section{Conclusions and perspectives}

Over the years, tensor networks have become highly efficient at simulating complex quantum systems, and researchers have only recently begun to explore their potential in other fields such as numerical simulations, optimization, and machine learning. In this work, we demonstrated how tensor networks can be highly effective in image processing, thanks to their remarkable compression capabilities. 

We showed how Fourier optics can be mapped to quantum mechanics, with qubits whose evolution is dictated by unitary operators that can be represented as matrix product operators, allowing efficient computations. In the future, we plan to extend this work to broader optical applications and explore algorithms for optical inverse problems—such as phase retrieval, denoising, and deconvolution—which have numerous applications in astronomy, earth observation, and microscopy.
\\
\\
\\
\textbf{Acknowledgements  } I am deeply grateful to Bernhard Jobst for discussions in the early stage of this paper, to all the contributors to the ITensor library ecosystem \cite{fishman2022itensor} and to one of the referee for its careful reading and constructive criticism. All simulations have been carried out on a 2015 macbook pro 16GB. The JPG picture used in this article will be made publicly available or can be obtained from the author upon request.

\bibliography{main}

\onecolumngrid
\newpage

\appendix

\section{Matrix product states and binary tensor trees}
\textcolor{black}{
In this appendix, we show explicitly how to construct MPS and tensor trees (figure below) that we used in the main text.
\begin{figure}[ht!]
\includegraphics[scale=0.3]{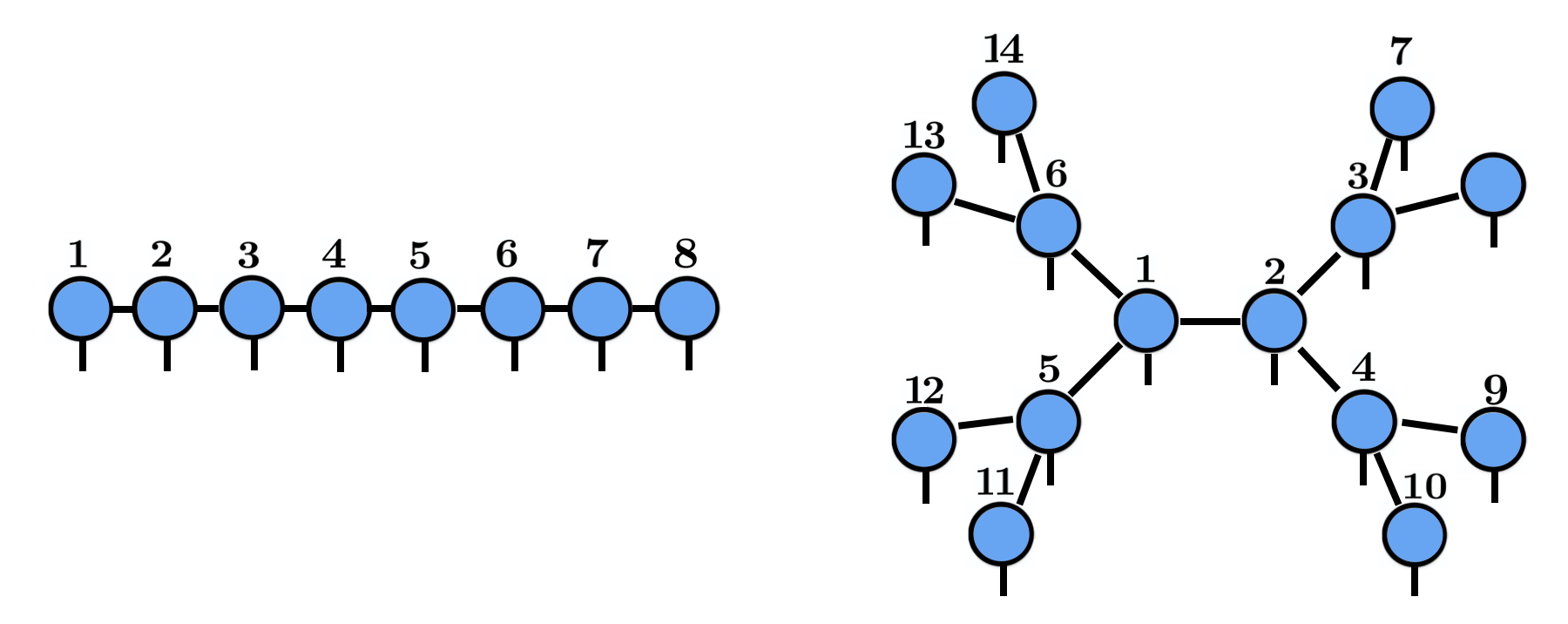} 
	\caption{Example of a matrix product state with $n=8$ and bulk binary tensor tree with $n=14$. The construction is shown below}
		\label{fig_plane}
\end{figure}
In both cases, the construction is based on successive singular value decompositions along the link of a network.  The full $n$-qubit state vector $|\psi\rangle\in\mathcal{H}$, whether we choose the FRQI or quantics encodings is 
\begin{equation}
|\psi\rangle = \sum_{s_1, s_2, ..., s_n \in \mathbb{Z}_2} \psi_{s_1 s_2 ... s_n}|s_1 s_2 ... s_n\rangle,
\end{equation}
 where the tensor $\psi_{s_1 s_2 \dots s_n}$ is represented by the purple $n$-leg object in \efig{MPS}.}

\subsection*{Matrix product states}

\textcolor{black}{
The matrix decomposition outlined below is a generalization of the singular value decomposition and can be used to decompose a generic tensor into constituent tensors of lower order. Every bond corresponds to a Schmidt decomposition of the Hilbert space $\mathcal{H}=\mathcal{H}_{[1,2,...,k]}\otimes\mathcal{H}_{[k+1,k+2,...,n]}$. The Schmidt decomposition across the cut can be written
\begin{equation}
|\psi\rangle
=\sum_{\alpha=1}^{\chi_k}
\lambda^{[k]}_{\alpha}
|\alpha\rangle_{[1,2,..,k]}
\otimes
|\alpha\rangle_{[k+1,k+2,..,n]}
\end{equation}
with orthonormality of the Schmidt bases generated by contracting left ($|\alpha'\rangle_{[1,2,..,k]}$) and right-canonical tensors ($|\alpha'\rangle_{[1,2,..,k]}$) such that $\langle \alpha | \alpha' \rangle_{[1,2,..,k]} = \delta_{\alpha\alpha'}$ and $\langle \alpha | \alpha' \rangle_{[k+1,k+2,..,n]} = \delta_{\alpha\alpha'}$.
Expressing the Schmidt basis states in the computational basis
\begin{align}
|\alpha\rangle_{[1,...,k]} &= \sum_{s_1,...,s_k} L^{[1..k]}_{s_1 ... s_k, \alpha} |s_1 ... s_k\rangle, \\
|\alpha\rangle_{[k+1,...,n]} &= \sum_{s_{k+1},...,s_n} R^{[k+1..n]}_{\alpha, s_{k+1} ... s_n} |s_{k+1} ... s_n\rangle.
\end{align}
Substituting back into the Schmidt decomposition gives
\begin{equation}
|\psi\rangle=
\sum_{s_1,\dots,s_n} \sum_{\alpha=1}^{\chi_k} 
\lambda^{[k]}_\alpha 
L^{[1..k]}_{s_1... s_k, \alpha} 
R^{[k+1..n]}_{\alpha, s_{k+1}... s_n}
|s_1 ... s_n\rangle.
\end{equation}
Factorizing left and right Schmidt states recursively
\begin{eqnarray}\nonumber
&&|\alpha\rangle_{[1,...,k]} = \sum_{\substack{\alpha_1,\dots,\alpha_{k-1} \\ s_1,...,s_k}}
A^{[1]}_{s_1, \alpha_1}
A^{[2]}_{\alpha_1, s_2, \alpha_2}..
A^{[k]}_{\alpha_{k-1}, s_k, \alpha}
|s_1... s_k\rangle, \nonumber\\
&&|\alpha\rangle_{[k+1,\dots,n]} = 
\sum_{\substack{\alpha_{k+1},...,\alpha_{n-1} \\ s_{k+1},...,s_n}}
A^{[k+1]}_{\alpha, s_{k+1}, \alpha_{k+1}}..
A^{[n]}_{\alpha_{n-1}, s_n}
|s_{k+1}...s_n\rangle,\nonumber
\end{eqnarray}
leads to the usual form of the factorization of components
\begin{equation}
\psi_{s_1 s_2 \dots s_n} =
\sum_{\alpha_1, \dots, \alpha_{n-1}}
A^{[1]}_{s_1, \alpha_1}
A^{[2]}_{\alpha_1, s_2, \alpha_2}..
A^{[n]}_{\alpha_{n-1}, s_n},
\end{equation}
which is represented diagrammatically by the blue object in \efig{MPS}. Depending on the scaling of $\chi_k$ with $n$, the MPS can efficiently represent $\psi_{s_1 s_2 \dots s_n}$}.

\subsection*{Binary tensor tree with physical sites in the bulk}

\textcolor{black}{Once fixed a binary tree topology, every SVD corresponds to a tree cut which can separate non-contiguous qubits in the original ordering $\{1,2,...,k\}|\{k+1,k+2,...,n\}$. We can decompose the full tensor into a binary tree of lower-order tensors using a sequence of Schmidt decompositions corresponding to the cuts defined by the edges of the tree. Every edge $e$ of the tree defines a bipartition of the Hilbert space
\begin{equation}
\mathcal{H} = \mathcal{H}_{\text{subtree}(e)} \otimes \mathcal{H}_{\text{rest}(e)}.
\end{equation}
Let us explicite the construction for the bulk tree (see \efig{fig_tree}). The Schmidt decomposition across the edge $e$ reads
\begin{equation}
|\psi\rangle
= \sum_{\alpha=1}^{\chi_e} \lambda^{(e)}_\alpha
|\alpha\rangle_{\text{subtree}(e)} \otimes |\alpha\rangle_{\text{rest}(e)},
\end{equation}
with orthonormality conditions
\begin{equation}
\langle \alpha | \alpha' \rangle_{\text{subtree}(e)} = \delta_{\alpha\alpha'}, \qquad
\langle \alpha | \alpha' \rangle_{\text{rest}(e)} = \delta_{\alpha\alpha'}.
\end{equation}
Expressing the Schmidt basis states in the computational basis, we have
\begin{align}
|\alpha\rangle_{\text{subtree}(e)} &= \sum_{\{s_v\}_{v \in \text{subtree}(e)}} 
L^{(e)}_{\{s_v\}, \alpha} \bigotimes_{v \in \text{subtree}(e)} |s_v\rangle\nonumber\\
|\alpha\rangle_{\text{rest}(e)} &= \sum_{\{s_w\}_{w \in \text{rest}(e)}} 
R^{(e)}_{\alpha, \{s_w\}}\bigotimes_{w \in \text{rest}(e)} |s_w\rangle.
\end{align}
Substituting back into the Schmidt decomposition gives
\begin{equation}
|\psi\rangle =
\sum_{s_1, \dots, s_n} \sum_{\alpha=1}^{\chi_e} 
\lambda^{(e)}_\alpha \,
L^{(e)}_{\{s_v\}, \alpha} \, R^{(e)}_{\alpha, \{s_w\}}
|s_1 s_2 \dots s_n\rangle.
\end{equation}
We can then factorize each subtree recursively into tensors associated with each node of the binary tree. For a node $v$ with children $c_1$ and $c_2$, we write
\begin{align}
|\alpha\rangle_{\text{subtree}(v)} = \sum_{\substack{s_v, \alpha_1, \alpha_2}} 
A^{(v)}_{s_v, \alpha_1, \alpha_2, \alpha} 
|\alpha_1\rangle_{\text{subtree}(c_1)} \otimes|\alpha_2\rangle_{\text{subtree}(c_2)} \otimes |s_v\rangle.
\end{align}
Finally, contracting all nodes of the tree produces the usual bulk binary tree decomposition of the tensor components
\begin{equation}
\psi_{s_1 s_2 \dots s_n} =
\sum_{\{\alpha_e\}} \prod_{v \in \text{nodes}} A^{(v)}_{s_v, \alpha_{v_1}, \alpha_{v_2}, \alpha_{v_\text{parent}}},
\end{equation}
where each $A^{(v)}$ has indices for the physical qubit at node $v$, the virtual bonds to its children, and the bond to its parent (if any). This is the direct generalization of the MPS canonical form to a binary tree geometry.}

\newpage

\section{Pauli string exact formula}

We derive an exact expansion of the function 
\begin{equation}
\cos\left( 2\pi \sum_{j=1}^n s_j 2^{-j} \right),
\end{equation}
where each $s_j \in \{0,1\}$ represents a binary digit. We express this function in terms of Pauli-$Z$ operators acting on an $n$-qubit Hilbert space and give an explicit formula for the expansion coefficients. We discuss properties of this expansion, its convergence in the thermodynamic limit, and implications for approximations in quantum many-body systems.
Consider an $n$-qubit computational basis labeled by bitstrings $\mathbf{s} = (s_1, s_2, \ldots, s_n)$ where each $s_j \in \{0,1\}$. Define the number
\begin{equation}
    x_{\mathbf{s}} = \sum_{j=1}^n s_j 2^{-j} \in [0,1).
\end{equation}
We want to express the function $f(\mathbf{s}) = \cos(2\pi x_{\mathbf{s}}) = \cos\left( 2\pi \sum_{j=1}^n s_j 2^{-j} \right)$
as an operator diagonal in the computational basis $\hat{F} = \sum_{\mathbf{s}} f(\mathbf{s}) \ket{\mathbf{s}}\bra{\mathbf{s}}$.
Recall the single-qubit Pauli-$Z$ operator acts as
$Z \ket{0} = \ket{0}$ and  $Z \ket{1} = -\ket{1}$.
For $n$ qubits, define operators
\begin{equation}
Z_j = I^{\otimes (j-1)} \otimes Z \otimes I^{\otimes (n-j)},
\end{equation}

acting non-trivially only on qubit $j$.
A convenient operator basis for diagonal operators on $n$ qubits is the set of all products
$\prod_{j \in S} Z_j$, where $S \subseteq \{1,2,\ldots,n\}$. For the empty set $S=\emptyset$, the product is the identity operator $I^{\otimes n}$.
These operators satisfy
\begin{equation}
\left( \prod_{j \in S} Z_j \right) \ket{\mathbf{s}} = (-1)^{\sum_{j \in S} s_j} \ket{\mathbf{s}}.
\end{equation}
Therefore, any diagonal operator can be expanded as
\begin{equation}
\hat{F} = \sum_{S \subseteq \{1,\ldots,n\}} c_S \prod_{j \in S} Z_j,
\end{equation}
with coefficients $c_S \in \mathbb{R}$. Because the operators $\prod_{j \in S} Z_j$ are diagonal and orthogonal under the Hilbert-Schmidt inner product, the coefficients can be extracted by
\begin{equation}
c_S = \frac{1}{2^n} \sum_{\mathbf{s} \in \{0,1\}^n} f(\mathbf{s}) (-1)^{\sum_{j \in S} s_j}.
\label{eq  coeff_formula}
\end{equation}
Inserting $f(\mathbf{s}) = \cos\left(2\pi \sum_{j=1}^n s_j 2^{-j}\right)$, we get
\begin{eqnarray}
c_S &=& \frac{1}{2^n} \sum_{\mathbf{s}} \cos\left( 2\pi \sum_{j=1}^n s_j 2^{-j} \right) (-1)^{\sum_{j \in S} s_j}\\
&=& \operatorname{Re}\left[ \frac{1}{2^n} \sum_{\mathbf{s}} e^{i 2\pi \sum_{j=1}^n s_j 2^{-j}} (-1)^{\sum_{j \in S} s_j} \right]\\
&=& \operatorname{Re} \left[ \frac{1}{2^n} \prod_{j=1}^n \sum_{s_j=0}^1 e^{i 2\pi s_j 2^{-j}} (-1)^{s_j \mathbf{1}_{j \in S}} \right].
\end{eqnarray}
Each sum over $s_j$ is
\begin{equation}
\sum_{s_j=0}^1 e^{i 2\pi s_j 2^{-j}} (-1)^{s_j \mathbf{1}_{j \in S}} = 1 + (-1)^{\mathbf{1}_{j \in S}} e^{i \theta_j},
\end{equation}
where we defined $\theta_j   = \frac{2\pi}{2^j}$.
Hence
\begin{equation}
\boxed{
c_S = \frac{1}{2^n} \operatorname{Re} \left[ \prod_{j=1}^n \left(1 + (-1)^{\mathbf{1}_{j \in S}} e^{i \theta_j} \right) \right].
}
\label{eq  final_coeff}
\end{equation}
The coefficient $c_S$ factorizes as a product of simple two-term factors. For $j \notin S$, the factor is $1 + e^{i \theta_j}$ and for $j \in S$, the factor is $1 - e^{i \theta_j}$. 
Since $\theta_j = \frac{2\pi}{2^j}$ decreases exponentially with $j$, the terms involving high $j$ contribute less.
The coefficients $c_S$ decay quickly in magnitude as $|S|$ increases or as the indices $j$ in $S$ grow.
As $n \to \infty$,
\begin{equation}
\hat{D} = \sum_{S \subseteq \mathbb{N}} c_S \prod_{j \in S} \hat{Z}_j,
\end{equation}
is an infinite series with absolutely convergent coefficients $c_S$ because
\begin{equation}
|1 \pm e^{i \theta_j}| \approx \sqrt{2 - 2\cos \theta_j} \sim \theta_j \sim 2^{-j} \to 0.
\end{equation}
This allows efficient approximations by truncation to low-order subsets $S$, making it useful for matrix product operator (MPO) approximations in quantum many-body systems.
We have derived a compact exact formula for the Pauli-$Z$ expansion coefficients of the function.

\subsection*{Unitary evolution}

We start from an operator given by a diagonal expansion in Pauli-$Z$ strings,
\begin{equation}
D=\sum_{S\subseteq\{1,\dots,n\}} c_S \, P_S,\qquad P_S\equiv\prod_{j\in S} Z_j,
\end{equation}
and derive the exact form of the time-evolution operator $U(t)=e^{-iHt}$. Because all $P_S$ commute and satisfy $P_S^2=I$, the unitary factorizes and each factor has a simple closed form. We give the derivation, show the action on the computational basis, and briefly discuss practical consequences.
Let $n$ be the number of (spin-1/2) sites. Denote by $Z_j$ the Pauli-$Z$ operator acting on site $j$ (identity elsewhere). 
Each $P_S$ is Hermitian and unitary
\begin{equation}
P_S^\dagger = P_S,\qquad P_S^2 = I,
\end{equation}
and different Pauli-$Z$ strings commute, because every $Z_j$ is diagonal and $Z_i Z_j = Z_j Z_i$ for all $i,j$
\begin{equation}
[P_S,P_{S'}]=0\quad\text{for all }S,S'.
\end{equation}
We want the time-evolution operator $U(t)=e^{-iDt}$.
Because the terms commute pairwise, the exponential of the sum factorizes exactly into the product of exponentials
\begin{equation}
U(t)  =  \prod_{S\subseteq\{1,\dots,n\}} e^{-i c_S t\, P_S}.
\end{equation}
Moreover, each factor $e^{-i\theta P_S}$ has a closed form because $P_S^2=I$. Using the power-series definition of the exponential and separating even/odd powers,
\begin{equation}
e^{-i\theta P_S} = \sum_{k=0}^\infty \frac{(-i\theta)^k}{k!} P_S^k
= \sum_{m=0}^\infty \frac{(-1)^m \theta^{2m}}{(2m)!} I
- i  \sum_{m=0}^\infty \frac{(-1)^m \theta^{2m+1}}{(2m+1)!} P_S,
\end{equation}
we obtain the elementary trigonometric form
\begin{equation}\label{eq  single-factor}
e^{-i\theta P_S}  =  \cos(\theta)\,I  -  i\sin(\theta)\,P_S.
\end{equation}
Applying this with $\theta=c_S t$ gives each factor explicitly. Therefore the exact evolution is
\begin{equation}\label{eq  U-product}
U(t)=\prod_{S}\big(\cos(c_S t)\,I - i\sin(c_S t)\,P_S\big).
\end{equation}
\subsection*{Exact derivation of the mean amplitude of Pauli-$Z$ coefficients by subset size}
We can rewrite $c_S$ by separating the product over indices in $S$ and not in $S$
\begin{align}
c_S &= \frac{1}{2^n} \operatorname{Re} \Bigg[ \prod_{j \in S} (1 - e^{i \theta_j}) \prod_{j \notin S} (1 + e^{i \theta_j}) \Bigg] \\
&= \frac{1}{2^n} \operatorname{Re} \Bigg[ \Big( \prod_{j>1} (1 + e^{i \theta_j}) \Big) \prod_{j \in S} \frac{1 - e^{i \theta_j}}{1 + e^{i \theta_j}} \Bigg],
\end{align}
for $j>1$. Only subsets $S$ that contain qubit $1$ ($C_S=0$ for $1\notin S$) give nonzero coefficients. Define 
\begin{equation}
x_j   = \frac{1 - e^{i \theta_j}}{1 + e^{i \theta_j}} = - i \tan(\theta_j / 2),
\end{equation}
so that
\begin{equation}
c_S = \frac{1}{2^n} \operatorname{Re} \left[ \left( \prod_{j>1} (1 + e^{i \theta_j}) \right) \prod_{j \in S} x_j \right].
\end{equation}
The mean coefficient over all strings of size $P$ is then
\begin{equation}
\overline{c}_P = \frac{1}{\binom{n}{P}} \sum_{|S|=P} c_S
= \frac{1}{2^n \binom{n}{P}} \operatorname{Re} \Bigg[ \Big( \prod_{j>1} (1 + e^{i \theta_j}) \Big) \sum_{|S|=P} \prod_{j \in S} x_j \Bigg].
\end{equation}
Recognizing the sum over products as the elementary symmetric polynomial $e_P(x_1,\dots,x_n)$, we obtain
\begin{eqnarray}
\overline{c}_P &=& \frac{1}{2^n \binom{n}{P}} \operatorname{Re} \left[ \left( \prod_{j>1} (1 + e^{i \theta_j}) \right) e_P(x_1, \dots, x_n) \right]\nonumber\\
&=&\frac{1}{2^n \binom{n}{P}} \operatorname{Re} \left[ \left(\frac{e^{i \pi \left(\frac{1}{2} - \frac{1}{2^n}\right)}}{\sin\left(\frac{\pi}{2^n}\right)}  \right) e_P(x_1, \dots, x_n) \right],
\end{eqnarray}
with $x_j = -i \tan(\theta_j/2)$. The \emph{elementary symmetric polynomial of degree $P$} in $n$ variables $x_1, \dots, x_n$ is defined as
\begin{equation}
e_P(x_1, \dots, x_n)   = \sum_{1 \le j_1 < j_2 < \dots < j_P \le n} x_{j_1} x_{j_2} \dots x_{j_P}, \quad e_0   = 1.
\end{equation}
For large $n$ and $P \ll n$, the $\theta_j$ for $j > P$ are small and contribute negligibly to the elementary symmetric polynomial. Therefore, the polynomial is dominated by the first $P$ terms
\begin{equation}
e_P(x_1, \dots, x_n) \sim \prod_{j=1}^P x_j = \prod_{j=1}^P (-i \tan(\theta_j/2)).
\end{equation}
Using $\theta_j / 2 = \pi / 2^j$, we have
\begin{equation}
\prod_{j=1}^P \tan(\theta_j/2) \sim \prod_{j=1}^P \frac{\pi}{2^j} = \pi^P \, 2^{-P(P+1)/2}.
\end{equation}
The product of cosines in the prefactor converges to a constant $C = \prod_{j=1}^\infty \cos(\pi / 2^j)$, and the division by $\binom{n}{P} \sim n^P / P!$ yields
\begin{equation}
\overline{c}_P \sim P! \, C \, \pi^P \, 2^{-P(P+1)/2} \, n^{-P}, \quad \text{for } P \ll n.
\end{equation}
Thus, the coefficients decay exponentially in $P(P+1)/2$ and polynomially in $n$, illustrating the rapid decrease of contributions from large strings.

\end{document}